\documentclass[aps,pra,groupedaddress,showpacs,twocolumn,superscriptaddress]{revtex4} 	

\usepackage{amsmath}
\usepackage{amssymb}
\usepackage{amsthm}
\usepackage{graphics}
\usepackage{pstricks}

\newcommand{\ad}{^\dagger }
\newcommand{\ket}[1]{|#1\rangle}     
\newcommand{\bra}[1]{\left\langle #1 \right|}     
\newcommand{\dyad}[2]{\ket{#1}\bra{#2}}           
\newcommand{\Tr}{{\rm Tr}}                        
\newcommand{\ii}{\mathrm{i}}					  

\newcommand{\vect}[1]{\mbox{\textbf{#1}}}         

\def\avg#1{\langle #1\rangle }

\def\dya#1{|#1\rangle \langle#1|}

\def\mat#1{\left(\begin{matrix}#1\end{matrix}\right)}
\def\om{\omega }
\def\ot{\otimes}

\def\QZ{\mbox{$Q$}}
\def\XZ{\mbox{$\XC_0$}}

\newcommand{\ZZ}{\mathbb{Z}}

\newcommand{\AC}{\mathcal{A}}
\newcommand{\BC}{\mathcal{B}}
\newcommand{\CC}{\mathcal{C}}

\newcommand{\EC}{\mathcal{E}}

\newcommand{\GC}{\mathcal{G}}
\newcommand{\HC}{\mathcal{H}}

\newcommand{\JC}{\mathcal{J}}
\newcommand{\KC}{\mathcal{K}}
\newcommand{\LC}{\mathcal{L}}

\newcommand{\OC}{\mathcal{O}}
\newcommand{\PC}{\mathcal{P}}
\newcommand{\QC}{\mathcal{Q}}
\newcommand{\RC}{\mathcal{R}}
\newcommand{\SC}{\mathcal{S}}

\newcommand{\WC}{\mathcal{W}}
\newcommand{\XC}{\mathcal{X}}

\newcommand{\GCbar}{\overline{\mathcal{G}}}

\newcommand{\pt}{B}   			
\newcommand{\ptc}{\bar\pt}	 	

\newtheorem{theorem}{Theorem}
\newtheorem{lemma}[theorem]{Lemma}

\begin{document}

\title{Location of quantum information in additive graph codes}
\author{Vlad Gheorghiu}
\email[Electronic address: ]{vgheorgh@andrew.cmu.edu}
\author{Shiang Yong Looi}
\email[Electronic address: ]{slooi@andrew.cmu.edu}
\thanks{The first two authors have contributed equally to this article}
\author{Robert B. Griffiths}
\affiliation{Department of Physics, Carnegie Mellon University, Pittsburgh,
Pennsylvania 15213, USA}

\date{Version of 8 December 2009}

\begin{abstract}
  The location of quantum information in various subsets of the qudit carriers
  of an additive graph code is discussed using a collection of operators on
  the coding space which form what we call the \emph{information group}.  It
  represents the input information through an encoding operation constructed
  as an explicit quantum circuit. Partial traces of these operators down to a
  particular subset of carriers provide an isomorphism of a subgroup of the
  information group, and this gives a precise characterization of what kinds
  of information they contain.  All carriers are assumed to have the same
  dimension $D$, an arbitrary integer greater than 1.
\end{abstract}

\maketitle


\tableofcontents

\section{Introduction\label{sct1}}


Quantum codes in which quantum information is redundantly encoded in a
collection of code carriers play an important role in quantum information, in
particular in systems for error correction and in schemes for quantum
communication \cite{PhysRevA.51.2738, PhysRevA.52.R2493, PhysRevLett.76.722, PhysRevA.54.1098}. They are a generalization of the classical codes well known and widely used in everyday communication systems \cite{MacWilliams:Sloane}. While for the latter it is fairly obvious where the information is located,
the quantum case is more complicated for two reasons.  First, a quantum
Hilbert space with its non-commuting operators is a more complex mathematical
structure than the strings of bits or other integers used in classical codes.
Second, the very concept of ``information'' is not easy to define in the
quantum case.  However, in certain cases one is able to make quite precise
statements.  Thus in the five qubit code \cite{PhysRevLett.77.198} that
encodes one qubit of information, none of the encoded information is present
in any two qubits taken by themselves, whereas all the information can be
recovered from any set of three qubits \cite{PhysRevA.71.042337}.


Similar precise statements can be made, as we shall see, in the case of an
\emph{additive graph code} on a collection of $n$ qudits which constitute the
\emph{carriers} of the code, provided each qudit has the same dimension $D$,
with $D$ some integer greater than one (not necessarily prime). It was shown
in \cite{PhysRevA.78.042303} that all additive graph codes are stabilizer
codes, and in \cite{quantph.0111080,quantph.0703112} that all stabilizer codes
are equivalent to graph codes for prime $D$. A detailed discussion of
non-binary quantum error correcting codes can be found in
\cite{quantph.9802007,IEEETransInfTheory.45.1827,PhysRevA.65.012308,
  PhysRevA.78.012306, PhysRevA.78.042303}.  The five qubit code just mentioned
is an example of a quantum code that is locally equivalent to an additive
graph code \cite{PhysRevA.65.012308}, and the information location has an
``all or nothing'' character.  In general the situation is more
interesting in that some subset of carriers may contain some but not all of
the encoded information, and what is present can be either ``classical'' or
``quantum,'' or a mixture of the two. Since many of the best codes currently known are additive graph
codes, identifying the location of information could prove useful when
utilizing codes for error correction, or designing new or better codes, or
codes that correct some types of errors more efficiently than others
\cite{PhysRevA.75.032345}. Our formalism can also be applied to study quantum secret sharing schemes employing graph states and can even handle a more general setting where there might be subsets that contain partial information and hence are neither authorized (contain the whole quantum secret) nor unauthorized (contain no information whatsoever about the secret).


Our approach to the problem of information location is algebraic, based upon
the fact that generalized Pauli operators on the Hilbert space of the carriers
form a group.  Subgroups of this group can be associated with different types
of information, and the information available in some subset of the carriers
can also be identified with, or is isomorphic to, an appropriate subgroup, as
indicated in the isomorphism theorem of Sec.~\ref{sct5}.  In the process of
deriving this theorem we go through a series of steps which amount to an
\emph{encoding procedure} that takes the initial quantum information and
places it in the coding subspace of the carrier Hilbert space.  These steps
can in turn be transformed into a set of quantum gates to produce an explicit
circuit that carries out the encoding.  This result, although somewhat
subsidiary to our main aims, is itself not without interest, and is an alternative to a previous scheme \cite{PhysRevA.65.012308} limited to prime $D$.


There have been some previous studies of quantum channels using an algebraic
approach similar to that employed here.  Those most closely related to our
work are by B\'{e}ny et al.\ \cite{PhysRevLett.98.100502,PhysRevA.76.042303}
(and see B\'{e}ny \cite{arxiv0907.4207}) and Blume-Kohout et al.\
\cite{PhysRevLett.100.030501}.  These authors have provided a set of very
general conditions under which an algebraic structure is preserved by a
channel.  In App.~\ref{apdxD} we show that our results fit within the
framework of a ``correctable algebra'' as defined in
\cite{PhysRevLett.98.100502,PhysRevA.76.042303,arxiv0907.4207}. See also the
remarks in Sec.~\ref{sct7}.


The remainder of this paper is organized as follows.  Some general comments
about types of quantum information and their connection with certain ideal
quantum channels are found in Sec.~\ref{sct2}. Section \ref{sct3} contains
definitions of the Pauli group and of some quantum gates used later in the
paper.  The formalism associated with additive graph codes as well as our
encoding strategy is in Sec.~\ref{sct4}; this along with some results on
partial traces leads to the fundamental isomorphism result in Sec.~\ref{sct5},
which also indicates some of its consequences for the types of information
discussed in Sec.~\ref{sct2}.  Section~\ref{sct6} contains various
applications to specific codes, for both qubit and qudit carriers.  Finally,
Sec.~\ref{sct7} contains a summary, conclusions, and some open questions.
Appendices~\ref{apdxA} and \ref{apdxB} contain longer proofs of theorems,
App.~\ref{apdxC} presents an efficient linear algebra based algorithm for
working out the results for any additive graph code, and App.~\ref{apdxD}
illustrates the connection with related work in \cite{PhysRevLett.98.100502}
and \cite{PhysRevA.76.042303}.

\section{Types of Information}
\label{sct2}


Both classical and quantum information theory have to do with statistical
correlations between properties of two or more systems, or properties of a
single system at two or more times.  In the classical case information is
always related to a possible set of physical properties that are distinct and
mutually exclusive---e.g., the voltage has one of a certain number of
values---with one and only one of these properties realized in a particular
system at a particular time.  For quantum systems it is useful to distinguish
different \emph{types} or \emph{species} of information
\cite{PhysRevA.76.062320}, each corresponding to a collection of mutually
distinct properties represented by a (projective) decomposition $\JC=\{J_j\}$
of the identity $I$ on the relevant Hilbert space $\HC$:
\begin{equation}
\label{eqn1}
  I = \sum_j J_j,\quad J_j = J_j^\dagger = J_j^2,
\quad J_j J_k = \delta_{jk} J_j.
\end{equation}
Any normal operator $M$ has a spectral representation of the form
\begin{equation}
\label{eqn2}
M = \sum_j \mu_j J_j,
\end{equation}
where the $\mu_j$ are its eigenvalues, and the decomposition $\{J_j\}$ is
uniquely specified by requiring $\mu_j\neq \mu_k$ when $j\neq k$. This means
one can sensibly speak about the type of information $\JC(M)$ \emph{associated
  with} a normal operator $M$.  When $M$ is Hermitian this is the kind of
information obtained by measuring $M$.


This terminology allows one to discuss the transmission of information through
a quantum channel in the following way.  Let $\EC$ be the
completely positive, trace preserving superoperator that maps the space of
operators $\LC(\HC)$ of the channel input onto the corresponding operator
space $\LC(\HC')$ of the channel output $\HC'$ (which may have a different
dimension from $\HC$).  Provided
\begin{equation}
  \EC(J_j) \EC(J_k) = 0 \text{ for } j\neq k,
\label{eqn3}
\end{equation}
for all the operators $\{J_j\}$ associated with a decomposition $\JC$ of the
$\HC$ identity, we shall say the channel is \emph{ideal} or \emph{noiseless}
for the $\JC$ species of information, or, equivalently, the $\JC$ type of
information is \emph{perfectly present} in the channel output $\HC'$.
Formally, each physical property $J_j$ at the input corresponds in a
one-to-one fashion to a unique property, the support of $\EC(J_j)$ (or the
corresponding projector) at the output.  Thus we have a quantum version of a
noiseless classical channel, a device for transmitting symbols, in this case
the label $j$ on $J_j$, from the input to the output by associating distinct
symbols with distinct physical properties---possibly a different collection of
properties at the output than at the input.


The opposite extreme from a noiseless channel is one in which $\EC(J_j)$ is
\emph{independent of $j$} up to a multiplicative constant.  In this case no
information of type $\JC$ is available at the channel output: the channel is
\emph{blocked}, or completely noisy; equivalently, the $\JC$ species of
information is \emph{absent} from the channel output.  Hereafter we shall
always use ``absent'' in the strong sense of ``completely absent'', and the
term \emph{present}, or \emph{partially present} for situations in which some
type of information is not (completely) absent but is also not perfectly
present: i.e., the channel is noisy but not completely blocked for this type
of information. 


In some cases all the projectors in $\{J_j\}$ will be of rank 1, onto pure
states, but in other cases some or all of them may be of higher rank, in which
case one may have a \emph{refinement} $\LC = \{L_l\}$ of $\{J_j\}$ such that
each projector $J_j$ is a sum of one or more projectors from the $\LC$
decomposition.  It is then clear that if the $\LC$ information is absent/perfectly present from/in the channel output the same is true of the $\JC$
information, but the converse need not hold.  Thus it may be that the coarse grained $\JC$
information is perfectly present, but no additional information is available
about the refinement.  A particularly simple situation, which we will
encounter later, is one in which the output $\HC'$ is itself a tensor product,
say $\HC'_1\ot\HC'_2$, $\JC$ a decomposition of $\HC'_1$, $\JC=\{J_j\otimes I\}$ and $\KC$ a decomposition of $\HC'_2$, $\KC=\{I\otimes K_k\}$. It can then be the case that the information associated with
the $\JC$ decomposition is perfectly present and that associated with the 
$\KC$ decomposition is (perfectly) absent from the channel output. 


Suppose $\JC = \{J_j\}$ and $\KC = \{K_k\}$ are two types of quantum
information defined on the same Hilbert space. The species $\JC$ and $\KC$ are
\emph{compatible} if all the projectors in $\JC$ commute with all the
projectors in $\KC$, in which case the distinct nonzero projectors in the
collection $\{J_j K_k\}$ provide a common refinement of the type discussed
above. Otherwise, if some projectors in one collection do not commute with
certain projectors in the other, $\JC$ and $\KC$ are \emph{incompatible} and
cannot be combined with each other. This is an example of the single framework
rule of consistent quantum reasoning, \cite{PhysRevA.54.2759} or Ch.~16 of
\cite{RBGriffiths:ConsistentQuantumTheory}.  The same channel may be ideal for
some $\JC$ and blocked for some $\KC$, or noisy for both but with different
amounts of noise.  From a quantum perspective, classical information theory is
only concerned with a single type of (quantum) information, or several
compatible types which possess a common refinement, whereas the task of
quantum, in contrast to classical, information theory is to analyze situations
where multiple incompatible types need to be considered.  


The term ``classical information'' when used in a quantum context can be
ambiguous or misleading.  Generally it is used when only a single type of
information, corresponding to a single decomposition of the identity, suffices
to describe what emerges from a channel, and other incompatible types can
therefore be ignored.  Even in such cases it is helpful to indicate explicitly
which decomposition of the identity is involved if that is not obvious from
the context. The contrasting term ``quantum information'' can then refer to
situations where two or more types of information corresponding to
incompatible decompositions are involved, and again it is helpful to be
explicit about what one has in mind if there is any danger of ambiguity.


An \emph{ideal quantum channel} is one in which there is an isometry $V$ from
$\HC$ to $\HC'$ such that 
\begin{equation}
  \EC(A) = VAV^\dagger
\label{eqn4}
\end{equation}
for every operator $A$ on $\HC$. In this case the superoperator $\EC$
preserves not only sums but also operator products:
\begin{equation}
  \EC(AB) = \EC(A) \EC(B).
\label{eqn5}
\end{equation}
Conversely, if \eqref{eqn5} holds for any pair of operators, one can show that
the quantum channel is ideal \cite{PhysRevLett.98.100502,PhysRevA.76.042303},
i.e. $\EC$ has the form \eqref{eqn4}.  As the isometry maps orthogonal
projectors to orthogonal projectors, \eqref{eqn3} will be satisfied for every
species of information, and we shall say that \emph{all} information is
perfectly present at the channel output.  The converse, that a channel which
is ideal for all species, or even for an appropriately chosen pair of
incompatible species is an ideal quantum channel, is also correct; see
\cite{PhysRevA.71.042337, PhysRevA.76.062320}.


The preservation of operator products, \eqref{eqn4}, can be a very useful tool
in checking for the presence or absence of various types of information in the
channel output, as we shall see in Sec.~\ref{sct5}.  When \eqref{eqn5} holds
for arbitrary $A$ and $B$ belonging to a particular decomposition of the
identity, this suffices to show that the channel is ideal for this
species. However, note that this sufficient condition is not necessary, since
\eqref{eqn3} could hold without the $\EC(A_j)$ being projectors, in which case
$\EC(A_j^2)$ is not mapped to $\EC(A_j)^2$.


We use the term \emph{ideal classical channel} for a type of information 
$\JC = \{J_j\}$ to refer to a situation where \eqref{eqn3} is satisfied and,
in addition, 
\begin{equation}
\label{eqn6} 
\EC(J_j A J_k) = 0 \text{ for } j\neq k,
\end{equation}
where $A$ is any operator on the input Hilbert space $\HC$. 
That is, not only is type $\JC$ perfectly transmitted, but all other types
are ``truncated'' relative to this type, in the notation of
\cite{PhysRevA.54.2759}.  

\section{Preliminary Remarks and Definitions}
\label{sct3}

\subsection{Generalized Pauli operators on $n$ qudits}
\label{sbsct3A}


We generalize Pauli operators to higher dimensional systems of arbitrary
dimension $D$ in the following way. The $X$ and $Z$ operators acting on a
single qudit are defined as
\begin{equation}
\label{eqn7}
Z=\sum_{j=0}^{D-1}\omega^j\dyad{j}{j},\quad X=\sum_{j=0}^{D-1}\dyad{j}{j+1},
\end{equation}
and satisfy
\begin{equation}
\label{eqn8}
X^D=Z^D=I,\quad XZ=\omega ZX,\quad \omega = \mathrm{e}^{2 \pi \ii /D},
\end{equation}
where \emph{the addition of integers is modulo $D$}, as will be 
assumed from now on. For a collection of $n$ qudits we use subscripts to
identify the corresponding Pauli operators: thus $Z_i$ and $X_i$ operate on
the space of qudit $i$. The Hilbert space of a single qudit is denoted by
$\HC$, and the Hilbert space of $n$ qudits by $\HC_n$,
respectively. Operators of the form
\begin{equation}
\label{eqn9}
\omega^{\lambda}X^{\vect{x}}Z^{\vect{z}} :=
\omega^{\lambda}X_1^{x_1}Z_1^{z_1}\otimes X_2^{x_2}Z_2^{z_2}\otimes\cdots
\otimes X_n^{x_n}Z_n^{z_n}
\end{equation} 
will be referred to as \emph{Pauli products}, where $\lambda$ is an integer
in $\ZZ_D$ and $\vect{x}$ and $\vect{z}$ are $n$-tuples in $\ZZ_D^n$, the
additive group of $n$-tuple integers mod $D$.  For a fixed $n$ the collection
of all possible Pauli products \eqref{eqn9} form a group under operator
multiplication, the \emph{Pauli group} $\PC_n$. If $p$ is a Pauli product,
then $p^D=I$ is the identity operator on $\HC_n$, and hence the order of any
element of $\PC_n$ is either $D$ or else an integer that divides $D$. While
$\PC_n$ is not abelian, it has the property that two elements \emph{commute up
  to a phase}: $p_1p_2 = \omega^{\lambda_{12}} p_2p_1$, with $\lambda_{12}$ an
integer in $\ZZ_D$ that depends on $p_1$ and $p_2$.


The collection of Pauli products with $\lambda=0$, i.e. a pre-factor of $1$, is
denoted by $\QC_n$ and forms an orthonormal basis of
$\LC(\HC_n)$, the Hilbert space of linear operators on
$\HC_n$, with respect to the inner product
\begin{equation}
\label{eqn10}
\frac{1}{D^n}\Tr[q_1^\dagger q_2]=\delta_{q_1,q_2}, \quad \forall q_1,q_2\in \QC_n.
\end{equation}
Note that $\QC_n$ is a \emph{projective group} or group up
to phases. There is a bijective map between $\QC_n$ and the quotient group
$\PC_n /\{\omega^{\lambda}{I}\}$ for $\lambda\in\ZZ_D$ where
$\{\omega^{\lambda}{I}\}$, the center of $\PC_n$, consists of phases
multiplying the identity operator on $n$ qudits.

\subsection{Generalization of qubit quantum gates to higher dimensions}
\label{sbsct3B}


In this subsection we define some one and two qudit gates generalizing
various qubit gates. The qudit generalization of the Hadamard gate is the
\emph{Fourier gate}
\begin{equation}\label{eqn11}
 \mathrm{F}:=\frac{1}{\sqrt{D}}\sum_{j=0}^{D-1}\omega^{jk}\dyad{j}{k}.
\end{equation}
For an invertible integer $q\in\ZZ_D$ (i.e. integer for which there exists $\bar q\in\ZZ_D$ such that $q \bar q \equiv 1 \bmod D$), we define a
\emph{multiplicative gate}
\begin{equation}\label{eqn12}
 \mathrm{S}_q:=\sum_{j=0}^{D-1}\dyad{j}{jq},
\end{equation}
where $qj$ means multiplication mod $D$. The requirement that $q$ be
invertible ensures that $\mathrm{S}_q$ is unitary; for a qubit
$\mathrm{S}_q$ is just the identity.


For two distinct qudits $a$ and $b$ we define the CNOT gate as
\begin{equation}
\label{eqn13}
 \mathrm{CNOT}_{ab}:=\sum_{j=0}^{D-1}\dyad{j}{j}_a\otimes X_b^j=\sum_{j,k=0}^{D-1}\dyad{j}{j}_a\otimes \dyad{k}{k+j}_b,
\end{equation}
the obvious generalization of the qubit Controlled-NOT, where $a$ labels the control qudit and $b$ labels the target qudit. Next the SWAP gate is defined as
\begin{equation}
\label{eqn14}
 \mathrm{SWAP}_{ab}:=\sum_{j,k=0}^{D-1}\dyad{k}{j}_a\otimes \dyad{j}{k}_b.
\end{equation}
It is easy to check that SWAP gate is hermitian and does indeed swap
qudits $a$ and $b$. Unlike the qubit case, the qudit SWAP gate is not a
product of three CNOT gates, but can be expressed in terms of CNOT gates and
Fourier gates as
\begin{equation}\label{eqn15}
 \mathrm{SWAP}_{ab}=\mathrm{CNOT}_{ab}(\mathrm{CNOT}_{ba})^{\dagger}\mathrm{CNOT}_{ab}(\mathrm{F}_a^2\otimes I_b),
\end{equation}
with 
\begin{equation}\label{eqn16}
(\mathrm{CNOT}_{ba})^{\dagger}=(\mathrm{CNOT}_{ba})^{D-1}=(I_a\otimes \mathrm{F}_b^2)\mathrm{CNOT}_{ba} (I_a\otimes \mathrm{F}_b^2). 
\end{equation}
Finally we define the generalized Controlled-phase or CP gate as
\begin{equation}
\label{eqn17}
\mathrm{CP}_{ab}=\sum_{j=0}^{D-1}\dyad{j}{j}_a\otimes Z^j_b=
\sum_{j,k=0}^{D-1}\omega^{jk}\dyad{j}{j}_a\otimes\dyad{k}{k}_b.
\end{equation}
The CP and CNOT gates are related by a local Fourier gate, similar to the qubit case
\begin{equation}\label{eqn18}
\mathrm{CNOT}_{ab}=(I_a\otimes \mathrm{F}_b) \mathrm{CP}_{ab} (I_a\otimes \mathrm{F}_b)^\dagger,
\end{equation}
since $\mathrm{F}$ maps $Z$ into $X$ under conjugation (see Table \ref{tbl1}).


The gates $\mathrm{F}$, $\mathrm{S}_q$, SWAP, CNOT and CP are unitary
operators that map Pauli operators to Pauli operators under conjugation, as
can be seen from Tables ~\ref{tbl1} and~\ref{tbl2}. They are elements of the
so called \emph{Clifford group} on $n$ qudits \cite{quantph.9802007,PhysRevA.71.042315}, the group of $n$-qudit unitary
operators that leaves $\PC_n$ invariant under conjugation, i.e. if $O$ is a
Clifford operator, then $\forall p\in\PC_n$, $OpO^\dagger\in\PC_n$. From
Tables~\ref{tbl1} and~\ref{tbl2} one can easily deduce the result of
conjugation by $\mathrm{F}$, $\mathrm{S}_q$, SWAP, CNOT and CP on \emph{any}
Pauli product. 

\begin{table}
\begin{tabular}{|l|l|l|}
\hline
Pauli operator    & $\mathrm{S}_q$ & $\mathrm{F}$  \\
\hline
\hline
$Z$ & $Z^{q}$ & $X$\\
\hline
$X$ & $X^{\bar q}$ & $Z^{D-1}$\\
\hline
\end{tabular}
\caption{The conjugation of Pauli operators by one-qudit gates $\mathrm{F}$ and $\mathrm{S}_q$ ($\bar q$ is the multiplicative inverse of $q$ mod $D$).}
\label{tbl1}
\end{table}

\begin{table}
\begin{tabular}{|l|l|l|l|}
\hline
Pauli product & $\mathrm{CNOT}_{ab}$ & $\mathrm{SWAP}_{ab}$ & $\mathrm{CP}_{ab}$\\
\hline
\hline
$I_a\otimes Z_b$ & $Z_a\otimes Z_b$ & $Z_a\otimes I_b$ & $I_a\otimes Z_b$\\ 
\hline
$Z_a\otimes I_b$ & $Z_a\otimes I_b$ & $I_a\otimes Z_b$ & $Z_a\otimes I_b$\\
\hline
$I_a\otimes X_b$ & $I_a\otimes X_b$ & $X_a\otimes I_b$ & $Z_a^{D-1}\otimes X_b$\\
\hline
$X_a\otimes I_b$ & $X_a\otimes X_b^{D-1}$ & $I_a\otimes{X}_b$ & $X_a\otimes Z_b^{D-1}$\\
\hline
\end{tabular}
\caption{
  The conjugation of Pauli products on qudits $a$ and $b$ by two-qudit 
  gates CNOT, SWAP and CP. For the CNOT gate, the first qudit $a$ is the
  control and the second qudit $b$ the target.
}
\label{tbl2}
\end{table}

\section{Graph States, Graph Codes and Related Operator Groups \label{sct4}}

\subsection{Graph states and graph codes\label{sbsct4A}} 


Let $G=(V,E)$ be a graph with $n$ vertices $V$, each corresponding to a qudit, and a collection $E$ of undirected edges connecting pairs of distinct vertices (no self loops). Two qudits can be joined by multiple edges, as long as the multiplicity does not exceed $D-1$. The graph $G$ is completely specified by the \emph{adjacency matrix} $\Gamma$, where the matrix element $\Gamma_{ab}$ represents the number of edges that connect vertex $a$ with vertex $b$. The \emph{graph state} 
\begin{equation}\label{eqn19}
\ket{G}=U\ket{G_0}=U\left(\ket{+}^{\otimes n}\right)
\end{equation}
is obtained by applying the unitary (Clifford) operator 
\begin{equation}
\label{eqn20}
U=\prod_{(a,b) \in E}\left(\mathrm{CP}_{ab}\right)^{\Gamma_{ab}},
\end{equation}
where each pair $(a,b)$ of vertices occurs only once in the product,
to the \emph{trivial graph state} 
\begin{equation}\label{eqn21}
\ket{G_0}:=\ket{+}^{\otimes n},
\end{equation}
with
\begin{equation}\label{eqn22}
\ket{+}:=\frac{1}{\sqrt{D}}\sum_{j=0}^{D-1}\ket{j}.
\end{equation}


Define $\SC^G$ to be the stabilizer of $\ket{G}$, the subgroup of operators from $\PC_n$ that leave $\ket{G}$ unchanged. The stabilizer $\SC^G_0$ of the trivial graph state $\ket{G_0}$ is simply the set of all $X$-type Pauli products with no additional phases,
\begin{equation}
\label{eqn23}
\SC^G_0=\left\{ X^{\vect{x}}: \vect{x}=(x_1,x_2,\ldots,x_n)\right\},
\end{equation}
where $x_j$ are arbitrary integers between $0$ and $D-1$.
Since $\ket{G}$ is related to $\ket{G_0}$ through a Clifford operator (see \eqref{eqn19} and \eqref{eqn20}), it follows at once that the stabilizer $\SC^G$ of $\ket{G}$ is related to the stabilizer $\SC_0^G$ of the trivial graph through the Clifford conjugation
\begin{equation}\label{eqn24}
\SC^G=U\SC^G_0 U^\dagger,
\end{equation}
with $U$ defined in \eqref{eqn20}.


A \emph{graph code} $C$ can be defined as the $K$-dimensional subspace $\HC_C$
of $\HC_n$ spanned by a collection of $K$ mutually orthogonal
codewords
\begin{equation}
\label{eqn25}
 \ket{\vect{c}_j} = 
Z^{\vect{c}_j}\ket{G},\quad j = 1,2,\ldots, K
\end{equation}
where 
\begin{equation}
\label{eqn26}
\vect{c}_j = (c_{j1}, c_{j2},\ldots, c_{jn})
\end{equation}
is for each $j$ an $n$-tuple in $\ZZ_D^n$.  The $c_{jk}$ notation suggests a
matrix $\vect{c}$ with $K$ rows and $n$ columns, of integers between $0$ and
$D-1$, and this is a very helpful perspective.  In this paper we are concerned
with \emph{additive} graph codes, meaning that the rows of this matrix form a
group under component-wise addition mod $D$, isomorphic to the abelian
\emph{coding group} $\CC$, of order $|\CC|=K$, of the operators
$Z^{\vect{c}_j}$ under multiplication.  We use $(\CC,\ket{G})$ to denote the
corresponding graph code.  For more details about graph states and graph codes
for arbitrary $D$, see \cite{PhysRevA.78.042303}.


Note that the codeword $(0,0,\ldots,0)$ is just the graph state $\ket{G}$, and
in the case of the trivial graph $\ket{G_0}$ this is the tensor product of
$\ket{+}$ states, \eqref{eqn21}, not the tensor product of $\ket{0}$ states
which the $n$-tuple notation $(0,0,\ldots,0)$ might suggest. Overlooking this
difference can lead to confusion through interchanging the role of $X$ and $Z$
operators, which is the reason for pointing it out here.

\subsection{The encoding problem\label{sbsct4B}}

A coding group $\CC$ can be used to create an additive code starting with any $n$ qudit
graph state, including the trivial graph $\ket{G_0}$, because the entangling
unitary $U$ commutes with $Z^{\vect{z}}$ for any $\vect{z}$; thus
\begin{equation}
\label{eqn27}
 \ket{\vect{c}_j} = Z^{\vect{c}_j}U\ket{G_0} = 
UZ^{\vect{c}_j}\ket{G_0} =U\ket{\vect{c}_j^0}
\end{equation}
where the $\ket{\vect{c}_j^0}$ span the code $(\CC,\ket{G_0})$. But in
addition the coding group $\CC$ is isomorphic, as explained below to a
\emph{trivial} code $\CC_0$,
\begin{equation}
\label{eqn28}
 \CC_0=\langle Z_1^{m_1}, Z_2^{m_2},\ldots, Z_k^{m_k}\rangle
\end{equation}
which is \emph{generated by}, i.e., includes all products of, the operators
inside the angular brackets $\langle$ $\rangle$.  Here $k$ is an integer less
than or equal to $n$, and each $m_j$ is $1$ or a larger integer that divides
$D$.
The simplest situation is the one in which each of the $m_j$ is equal to 1, in
which case $\CC_0$ is nothing but the group, of order $D^k$, of products of
$Z$ operators to any power less than $D$ on the first $k$ qudits.  One can
think of these qudits as comprising the input system through which information
enters the code, while the remaining $n-k$ qudits, each initially in a
$\ket{+}$ state, form the ancillary system for the encoding operation.  


If, however, one of the $m_j$ is greater than 1, the corresponding generator
$Z_j^{m_j}$ is of order
\begin{equation}
\label{eqn29}  
d_j = D/m_j,
\end{equation}
and represents a qudit of dimensionality $d_j$ rather than $D$.  Thus for
example, if $D=6$ and $m_1=2$, applying $Z_1^2$ and its powers to $\ket{+}$
will produce three orthogonal states corresponding to a qutrit, $d_1=3$.
(Identifying operators $Z$ and $X$ on these three states which satisfy
\eqref{eqn8} with $D=3$ is not altogether trivial, and is worked out in
Sec.~\ref{sbsct4C} below.)
In general one can think of the group $\CC_0$ in \eqref{eqn28} as
associated with a collection of $k$ qudits, the $j$'th qudit having dimension
$d_j$, and therefore the collection as a whole a dimension of $K=d_1d_2\cdots
d_k$, equal to that of the graph code.  If one thinks of the information to be
encoded as initially present in these $k$ qudits, the encoding problem is how
to map them in an appropriate way into the coding subspace $\HC$ of the $n$
($D$-dimensional) carriers.  


We address this by first considering
the connection between $\CC$ and $\CC_0$ in a simple 
example with $n=3$, $D=6$, and 
\begin{equation}
\label{eqn30}
\CC=\left\langle  Z_1^4Z_2^3Z_3^3, Z_2^3Z_3^3 \right\rangle,
\end{equation}
a coding group of order 6. The two generators in \eqref{eqn30} correspond, in
the notation introduced in \eqref{eqn26}, to the rows of the $2\times 3$
matrix
\begin{equation}
\label{eqn31}
\vect{f} =\mat{ 
 4 & 3 & 3\\
 0 & 3 & 3}.
\end{equation}
By adding rows or multiplying them by constants mod $D$ one can create 4
additional rows which together with those in \eqref{eqn31} constitute the
$6\times 3$ $\vect{c}$  matrix.


Through a sequence of elementary operations mod $D$---a) interchanging of
rows/columns, b) multiplication of a row/column by an \emph{invertible} integer, c)
addition of any multiple of a row/column to a \emph{different} row/column---a
matrix such as $\vect{f}$ can be converted to the Smith normal form 
\cite{Newman:IntegralMatrices,Storjohann96nearoptimal}
\begin{equation}
  \vect{s} = \vect{v}\cdot\vect{f}\cdot\vect{w},
\label{eqn32}
\end{equation}
where $\vect{v}$ and $\vect{w}$ are invertible (in the mod $D$ sense) square
matrices, and $\vect{s}$ is a diagonal rectangular matrix, as in
\eqref{eqn33}. It is proved in \cite{Storjohann96nearoptimal} that a $K\times n$ matrix can be reduced to the Smith form in only $\OC(K^{\theta-1}n)$ operations from $\ZZ_D$, where $\theta$ is the exponent for matrix multiplication over the ring $\ZZ_D$, i.e. two $m\times m$ matrices over $\ZZ_D$ can be multiplied in $\OC(m^{\theta})$ operations from $\ZZ_D$. Using standard matrix multiplication $\theta=3$, but better algorithms \cite{CoppersmithWinograd} allow for $\theta=2.38$. 


For the example above, the sequence
\begin{equation}
\label{eqn33}
  \mat{ 4 & 3 & 3\\ 0 & 3 & 3} \rightarrow 
  \mat{ 4 & 0 & 0\\ 0 & 3 & 3} \rightarrow 
  \mat{ 4 & 0 & 0\\ 0 & 3 & 0} \rightarrow 
  \mat{ 2 & 0 & 0\\ 0 & 3 & 0} =\vect{s}
\end{equation} 
proceeds by adding the second row of $\vect{f}$ to the first (mod 6), then the
second column to the third column, and finally multiplying the first row by 5
(which is invertible mod 6).  The final step is needed so that the diagonal
elements divide $D$: $m_1=2$, $m_2=3$, so that $d_1=3$ and $d_2=2$. Thus we
arrive at the trivial coding group
\begin{equation}
\label{eqn34}
 \CC_0=\left\langle Z_1^2, Z_2^3 \right\rangle,
\end{equation}
isomorphic to $\CC$ in \eqref{eqn30}.

\begin{figure}
\includegraphics{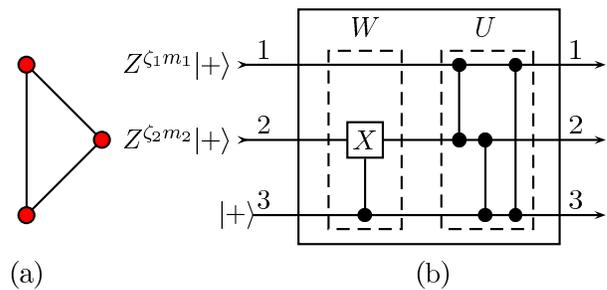}
\caption{(a) The graph state used in the example; (b) The encoding circuit: the input states ${Z_1^{\zeta_1 m_1}Z_2^{\zeta_2 m_2}\ket{++}}$ that correspond to the trivial code $\CC_0$ are mapped by
$W$ to $\CC$, then $U$ entangles the qudits. Here $m_1=2$, $m_2=3$ and $\zeta_j$ are integers such that $0\leqslant \zeta_j\leqslant d_j-1$, with $d_1=3$, $d_2=2$.}
\label{encoding}
\end{figure}

Since the procedure for reducing a matrix to Smith normal form is quite
general, the procedure illustrated in this example can be applied to
any coding group $\CC$, as defined following \eqref{eqn26}, to find a
corresponding trivial coding group $\CC_0$. 
The row operations change the collection of generators but not the coding
group that they generate; i.e., the final collection of $K$ rows is the same.
The column operations, on the other hand, produce a different, but isomorphic,
coding group, and one can think of these as realized by a unitary operator $W$
which is a product of various SWAP, CNOT and $\mathrm{S}_q$ gates, so that 
\begin{equation}
\label{eqn35}
\CC=W\CC_0 W^\dagger,
\end{equation}
that is, conjugation by $W$ maps each operator in $\CC_0$ to its counterpart
in $\CC$.  In our example, $W=\mathrm{CNOT}_{32}$ is the only column
operation, the second arrow in \eqref{eqn33}, and represents the first step
in the encoding circuit for this example, Fig.~\ref{encoding}(b). It is left
as an exercise to check that this relates the generators in \eqref{eqn30} and
\eqref{eqn34} through \eqref{eqn35}.  Table~\ref{tbl3} indicates how
different matrix column operations are related to the corresponding gates in
the encoding circuit. 

\begin{table}
\begin{tabular}{|l|c|}
\hline
Matrix operation in $\ZZ_D$ & Clifford conjugation\\
\hline
\hline
Interchange of columns $a$ and $b$    & $\mathrm{SWAP}_{ab}$\\
\hline
Multiplication of column $a$          & $\mathrm{S}_q$ on qudit $a$\\
by invertible integer $q$             & \\
\hline
Addition of m times column $b$ to             & $(\mathrm{CNOT}_{ab})^m$ \\
column $a$             &  \\
\hline
\end{tabular}
\caption{The correspondence between matrix column operations in $\ZZ_D$ and 
conjugation by Clifford gates. For the CNOT gate, the first qudit $a$ is the control
and the second qudit $b$ the target.}
\label{tbl3}
\end{table}


The overall encoding operation 
\begin{equation}
\label{eqn36}
\ket{\vect{c}_j}=UW\ket{\vect{c}_j^{0}}
\end{equation}
starting with the trivial code on the trivial graph $(\CC_0,\ket{G_0})$  and
ending with the desired code $(\CC,\ket{G})$ is shown for our example in
Fig.~\ref{encoding}(b) for the case of a graph indicated in (a) in this
figure.  It is important to notice that both $W$ and $U$, and therefore their
product, are Clifford operators, unitaries that under conjugacy map Pauli products to Pauli products.  This follows from the fact
that the gates in Table~\ref{tbl3} are Clifford gates, and will allow us in
what follows to extend arguments that are relatively straightforward for
trivial codes on trivial graphs to more general additive graph codes.  

\subsection{The information group}
\label{sbsct4C}


In this section we define the \emph{information group} that plays a central
role in the isomorphism theorem in Sec.~\ref{sct5} below.  The basic strategy
is most easily understood in terms of $C_0 = (\CC_0,\ket{G_0})$, the trivial
code on the trivial graph.  However, because the overall encoding map $UW$ in
\eqref{eqn36} is a Clifford operation mapping Pauli products to Pauli
products, various results that apply to $C_0$ can be immediately translated
to the general graph code $C = (\CC,\ket{G})$ we are interested in, and for
this reason most of the formulas valid for both are written in the form
valid for $C$ even if the derivations are based on $C_0$.

The pointwise stabilizer%
\footnote{Also called the ``fixer'' or ``fixator''.  It is important to
  distinguish this subgroup from the group theoretical notion of the
  stabilizer of the coding space in the sense of the
  subgroup of $\PC_n$ that maps the coding space onto itself without necessarily
  leaving the individual vectors fixed.  As we shall not employ the latter, it
  should cause no confusion if we hereafter follow the usual convention in
  quantum codes and omit ``pointwise,'' even though retaining it would add
  some precision.}%
\ of $C_0$, the subgroup of operators from $\PC_n$ that leave every codeword
$\ket{\vect{c}^0_j}$ unchanged, is given by
\begin{equation}
\label{eqn37}
\SC_0=\left\{X^{\vect{x}}:\vect{x}
 =(\eta_1 d_1,\eta_2 d_2,\ldots,\eta_k d_k,x_{k+1},\ldots,x_n)
\right\},
\end{equation}
where the $d_j$ are defined in \eqref{eqn29}, $\eta_j$ is any integer between
0 and $m_j-1$, and the $x_j$ for $j>k$ are arbitrary integers between 0 and
$D-1$.  That this is correct can be seen as follows.  First, Pauli products
belonging to $\SC_0$ cannot contain $Z_j$ operators, for such operators map
each codeword onto an orthogonal state. On the other
hand, every $X_j^{x_j}$ leaves $\ket{G_0}$, \eqref{eqn21}, unchanged, so it
belongs to $\SC_0$ if and only if it commutes with $Z_j^{m_j}$, which means
$x_jm_j$ must be a multiple of $D$, or $x_j$ a multiple of $d_j$, see
\eqref{eqn29}.  Thus elements of $\SC_0$ commute with elements of $\CC_0$,
\eqref{eqn28}.  Since its operators cannot alter the phases of the codewords,
no additional factors of $\om^\lambda$ are allowed, and thus $\SC_0$ is given
by \eqref{eqn37}.  The stabilizer of the (nontrivial) code $C$ is then the
isomorphic group $\SC$ obtained using the unitary $UW$ of \eqref{eqn36}:
\begin{equation}
\label{eqn38}
\SC = (UW)\SC_0(UW)^\dagger \equiv \{ (UW) s (UW)^\dag \ : s\in \SC_0 \}, 
\end{equation}
a collection of Pauli products because the unitary $UW$, as remarked earlier,
is a Clifford unitary.  The order of $\SC_0$, and thus of $\SC$, is given by
\begin{equation}
\label{eqn39}
|\SC|=D^{n-k}\prod_{j=1}^k{m_j}=
\frac{D^n}{\prod_{j=1}^kd_j}=\frac{D^n}{|\CC|}=\frac{D^n}{K}.
\end{equation}

Next define the subgroup $\WC$ of $\PC_n$
\begin{equation}
\label{eqn40}
\WC=\langle \SC^G, \CC\rangle
\end{equation}
generated by operators belonging to the stabilizer $\SC^G$ of the graph state
or to the coding group $\CC$, and denote it by $\WC_0=\langle \SC^G_0,
\CC_0^{}\rangle$ in the case of the trivial code. The elements of $\SC_0$
commute with those of $\SC_0^G$ (both are abelian and the former is a subgroup
of the latter), and also with those of $\CC_0$, as noted above.  As group
properties are preserved under the $UW$ map, as in \eqref{eqn38}, we conclude
that all elements in $\SC$ commute with those in $\WC$, even though $\WC$ is
not (in general) abelian, and hence $\SC$ is a normal subgroup of $\WC$. Now
define the \emph{abstract information group} as the quotient group
\begin{equation}
\label{eqn41}
\GCbar=\WC/\SC=\langle \SC^G, \CC\rangle/\SC
\end{equation}
consisting of cosets of $\SC$, written as $g\SC$ or $\SC g$ for $g$ in $\WC$.
Note that because any element $g$ of $\WC$ is a Pauli product, $g^D=I$ is the
identity, and the order of $g$ is either $D$ or an integer that divides $D$. 
Consequently the order of any element of $\GCbar$ is also $D$ or an integer
that divides $D$. 

To understand the significance of $\GCbar$ consider a
trivial code on a single qudit, with 
\begin{equation}
\label{eqn42} 
\CC_0 = \avg{Z_1^{m_1}},\quad \SC_0^G = \avg{X_1},\quad 
\SC_0 = \avg{X_1^{d_1}}
\end{equation}
The elements of $\GCbar_0$ can be worked out using its identity $\bar I$ and
the generators $\bar X$ and $\bar Z$:
\begin{align}
\label{eqn43} 
 \bar I &= \SC_0 = \{I_1, X_1^{d_1}, X_1^{2d_1},\ldots \}
\notag\\
 \bar X &= X_1\SC_0 = \{X_1, X_1^{d_1+1}, X_1^{2d_1+1},\ldots \}
\notag\\
 \bar Z &= Z_1^{m_1}\SC_0 = \{Z_1^{m_1}, Z_1^{m_1} X_1^{d_1},\ldots \}.
\end{align}
It is evident that the cosets $\bar X$, $\bar X^2 =
X_1^2\SC_0$ and so forth up to $\bar X^{d_1-1}$ are distinct, whereas
$\bar X^{d_1}=\bar I = \SC_0$.  The same is true for powers of $\bar Z$.
Furthermore,
\begin{equation}
\label{eqn44}
\bar X\bar Z = X_1 Z_1^{m_1}\SC_0 = \omega^{m_1} Z_1^{m_1}  X_1\SC_0 = 
 \bar\omega \bar Z\bar X,
\end{equation}
with $\bar\omega = \omega^{m_1} = \mathrm{e}^{2\pi \mathrm{i}/d_1}$.  Thus $\GCbar_0$ is 
generated by operators $\bar X$ and $\bar Z$ that satisfy \eqref{eqn8}
with $D$ replaced by $d_1$, which is to say the corresponding group is what
one would expect for a qudit of dimension $d_1$.  The same  argument
extends easily to the trivial code on $k$ carriers produced by $\CC_0$, see
\eqref{eqn28}: $\GCbar_0$ is isomorphic to the group of Pauli products on a
set of qudits of dimension $d_1, d_2,\ldots, d_k$.  The same
structure is inherited by the abstract information group $\GCbar$ for 
the code $C=(\CC,\ket{G})$ obtained by applying the $UW$ map as
in \eqref{eqn38}.

The abstract information group $\GCbar$ is isomorphic to the \emph{information
  group} $\GC$ of \emph{information operators} acting on the coding space $\HC_C$ and defined in
the following way.  Its identity is the operator
\begin{equation}
\label{eqn45}
 P = |\SC|^{-1}\Sigma(\SC) =  |\SC|^{-1}\sum_{s\in\SC} s,
\end{equation}
where $\Sigma(\AC)$ denotes the sum of the operators that make up a collection
$\AC$.  In fact, $P$ is just the projector onto $\HC_C$, as
can be seen as follows. Since $\SC$ is a group, $P^2=P$; and since a group
contains the inverse of every element, and $s\in \SC$ is unitary (a Pauli
product),  $P^\dagger = P$.  These two conditions mean that $P$ is a projector
onto some subspace of $\HC_n$.  Since $\SC$ is the (pointwise) stabilizer of
the coding space each $s$ in $\SC$ maps a codeword onto itself, and
thus $P$ maps each codeword to itself. Consequently, all the codewords lie in
the space onto which $P$ projects.  Finally, the rank of $P$ is
\begin{equation}
\label{eqn46}
\Tr[P] = D^n/|\SC| = |\CC| = K
\end{equation}
(see \eqref{eqn39}), since the trace of every $s$ in $\SC$ is zero except for
the identity with trace $D^n$. (Note that while $\PC_n$ contains the identity
multiplied by various phases, only the identity operator occurs in $\SC$.)
Therefore $P$ projects onto $\HC_C$, and is given by the formula
\begin{equation}
\label{eqn47}
P= \sum_{j=1}^{K} \dya{\vect{c}_j}. 
\end{equation}

The other information operators making up the information group
$\GC=\{\hat g\}$ are formed in a similar way from the different cosets
making up $\WC/\SC$:
\begin{equation}
\label{eqn48} 
 \hat g = |\SC|^{-1}\Sigma(g\SC) =  g P = P g P = P\hat g P.
\end{equation}
That is, for each coset form the corresponding sum of operators and divide by
the order of the stabilizer $\SC$.  The second and third equalities in
\eqref{eqn48} reflect the fact that the product of the cosets $\SC$ and $g\SC$
in either order is $g\SC$, which is to say $P$ forms the group identity of
$\GC$. They also tell us that the operators that make up $\GC$ act only on the
coding space, mapping $\HC_C$ onto itself, and give zero when applied to any
element of $\HC_n$ in the orthogonal complement of $\HC_C$.  Because $\SC$ is
a normal subgroup of $\WC$, products of operators of the form \eqref{eqn48}
mirror the products of the corresponding cosets, so the map from the abstract
$\GCbar$ to the group $\GC$ is a homomorphism.  That it is actually an
isomorphism is a consequence of the following, proved in App.~\ref{apdxA}:

\begin{lemma}\label{thm1}

Let $\RC$ be a linearly independent collection of Pauli product operators that
form a subgroup of $\PC_n$, and for a Pauli product $p$ let $p\RC=\{pr: r\in
\RC\}$. Then

i) The operators in $p\RC$ are linearly independent.

ii) If $p$ and $q$ are two Pauli products, one or the other of the
following two mutually exclusive possibilities obtains:

$\alpha$) 
\begin{equation}
\label{eqn49} 
p\RC = \mathrm{e}^{\ii\phi} q\RC
\end{equation}
in the sense that each operator in $p\RC$ is equal to $\mathrm{e}^{\ii\phi}$ times an
operator in $q\RC$

$\beta$)
The union $p\RC\cup q\RC$ is a collection of $2|\RC|$ linearly independent
operators. 

\end{lemma}

Since the collection of Pauli products $\QC_n$ with fixed phase forms a basis
of $\LC(\HC_n)$, a collection of Pauli products can be linearly
\emph{dependent} if and only if it contains both an operator and that operator
multiplied by some phase.  As the (pointwise) stabilizer $\SC$ leaves each
codeword unchanged, the corresponding operators are linearly independent, and
the lemma tells us that distinct cosets $g\SC\neq h\SC$ give rise to distinct
operators $\hat g\neq\hat h$.  Either $g\SC = \mathrm{e}^{\ii\phi} h\SC$, in which
case $\hat g = \mathrm{e}^{\ii\phi} \hat h \neq \hat h$ (since if $\mathrm{e}^{\ii\phi}=1$ the cosets are identical). Or else the $g\SC$ operators are linearly independent
of the $h\SC$ operators, and therefore $\hat g$ and $\hat h$ are linearly
independent. Consequently the homomorphic map from $\GCbar$ to $\GC$ is a
bijection, and the two groups are isomorphic.

The single qudit example considered in \eqref{eqn42} provides an example of how
$\GCbar$ and $\GC$ are related.  In this case the projector
\begin{equation}
\label{eqn50} 
P_0 = (1/m_1) (I_1 + X_1^{d_1} + \cdots)
\end{equation}
projects onto the subspace spanned by $\ket{+},Z_1^{m_1}\ket{+}, Z_1^{2
  m_1}\ket{+},\ldots$. While each of the operators that make up a coset such as
$\bar X$ in \eqref{eqn43} is unitary, their sum, an operator times $P_0$, is
no longer unitary, though when properly normalized acts as a unitary on the
subspace onto which $P_0$ projects. That the different sums of operators
making up the different cosets are distinct is in this case evident from
inspection without the need to invoke Lemma~\ref{thm1}.

Let us summarize the main results of this subsection.  For an additive graph
code $C$ we have defined the information group $\GC$ of operators acting on
the coding subspace $\HC_C$, whose group identity is the projector $P$ onto $\HC_C$.  
It is isomorphic to the group of Pauli products acting on a tensor product of
qudits of dimensions $d_1$, $d_2$, \dots, $d_k$, which can be thought of as the
input to the code, see Sec.~\ref{sbsct4B}.  Each element $\hat g$ of $\GC$ is
of the form $P\hat g P$, so as an operator on $\HC_n$ it commutes with $P$ and
yields zero when applied to any vector in the orthogonal complement of
$\HC_C$. The dimension of $\HC_C$ is $K=d_1 d_2\cdots d_k$, the size of the
code, and hence the elements of $\GC$ span the space of linear operators
$\LC(\HC_C)$ on $\HC_C$.

\section{Subsets of Carriers and the Isomorphism Theorem}
\label{sct5}

\subsection{Subsets of carriers}
\label{sbsct5A}

Before stating the isomorphism theorem, which is the principal technical
result of this paper, let us review some facts established in Sec.~\ref{sct4}.
The additive graph code $(\CC,\ket{G})$ we are interested in can be thought of
as arising from an encoding isometry that carries the channel input onto a
subspace $\HC_C$ of the $n$-qudit carrier space $\HC_n$, as in Fig~\ref{encoding}.  
This isometry, as explained in Sec.~\ref{sct2} in connection with \eqref{eqn4}, constitutes a perfect quantum channel, and thus all the information of interest can be said
to be located in the $\HC_C$ subspace, where it is represented by the
information group $\GC$, a multiplicative group of operators for which the
projector $P$ on $\HC_C$ is the group identity, and which as a group is
isomorphic to the abstract information group $\GCbar$ defined in
\eqref{eqn41}.

We are interested in what kinds of information are available in some subset
$\pt$ of the carriers, where $\ptc$ denotes the complementary set.  For this
purpose it is natural to consider the partial traces over $\ptc$, i.e., the
traces down to the Hilbert space $\HC_\pt$, of the form
\begin{equation}
\label{eqn51}
 g_\pt =  N^{-1} \Tr_{\ptc}[\hat g],
\end{equation}
where $\hat g$ is an element of the information group $\GC$, and the positive
constant $N$ is defined in \eqref{eqn58} below.  In those cases in which
$g_{\pt}=0$ the $\JC(\hat g)$ information has disappeared and is not available in the subset $B$, so we shall be interested in those $\hat g$ for which the partial trace does not vanish, that is to say in the elements of the \emph{subset information group}
\begin{equation}
\label{eqn52}
\GC^{\pt}=\left\{ \hat g \in \GC: \Tr_{\ptc}[\hat g]\neq 0\right\}.
\end{equation}
We show below that $\GC^{\pt}$ is a subgroup of $\GC$, thus justifying its name,
and that it is isomorphic to the group $\GC_{\pt}$ of nonzero operators of the form
$g_{\pt}$ defined in \eqref{eqn51}.  To actually determine which elements of $\GC$
belong to $\GC^\pt$ one needs to take partial traces of the $\hat g \in \GC$
to see which of them do not trace down to
zero. In App.~\ref{apdxC} we present an efficient linear algebra algorithm
based on solving systems of linear equations mod $D$ that can find $\GC^{\pt}$
in $\mathcal{O}(K^2n^\theta)$ operations from $\ZZ_D$ where $\theta$ is defined
in Sec. \ref{sbsct4B}.

If an operator $A$ on the full Hilbert space $\HC_n$ of the $n$ carriers
can be written as a tensor product of an operator on $\HC_\pt$ times the
identity operator $I_{\ptc}$ on $\HC_{\ptc}$ we shall say that $A$ is
\emph{based in $\pt$}. Let $\BC$ be the collection of all operators on $\HC_n$
that are based in $\pt$.  Obviously, $\BC$ is closed under sums, products,
and scalar multiplication. In addition the partial trace $\Tr_{\ptc}[A]$ of an
operator $A$ in $\BC$ is ``essentially the same'' operator, apart from
normalization in the sense that
\begin{equation}
\label{eqn53} 
A = D^{-|\ptc|}\cdot \Tr_{\ptc}[A] \otimes I_{\ptc}.
\end{equation}
If $A\notin\BC$ is a Pauli product, then its partial trace over $\ptc$
vanishes, since $\Tr[X]$ and $\Tr[Z]$ and their powers (when not equal to
$I$) are zero.  Consequently the partial trace over $\ptc$ of $\Sigma(g\SC)$ in
\eqref{eqn48} is the same as the partial trace of $\Sigma[(g\SC)\cap\BC]$,
which suggests that it is useful to consider the properties of collections of
Pauli operators of the form $(g\SC)\cap\BC$ with $g$ an element of $\WC$.
The following result, proved in App.~\ref{apdxA}, turns out to be useful.

\begin{lemma}\label{thm2}
Let $g,h$ be two arbitrary elements of $\WC$, and $\BC$ the collection of operators with
base in $\pt$.  

i) The set $(g\SC)\cap\BC$ is empty if and only if $(g^{-1}\SC)\cap\BC$ is
empty. 

ii) Every nonempty set of the form $(g\SC)\cap\BC$ contains precisely 
\begin{equation}
\label{eqn54} M = |\SC\cap \BC| \geq 1
\end{equation}
elements.

iii) Two nonempty sets $(g\SC)\cap\BC$ and $(h\SC)\cap\BC$ are either
identical, which means $g\SC = h\SC$ and $\Sigma[(g\SC)\cap\BC] =
\Sigma[(h\SC)\cap\BC]$, or  else they have no elements in common
and the operators $\Sigma[(g\SC)\cap\BC]$ and $\Sigma[(h\SC)\cap\BC]$
are distinct.

iv) If both $(g\SC)\cap\BC$ and $(h\SC)\cap\BC$ are nonempty, their product
as sets, including multiplicity, is given by
\begin{equation}
\label{eqn55}
[ (g\SC)\cap\BC]\cdot [(h\SC)\cap\BC] = M[ (gh\SC)\cap\BC].
\end{equation}

\end{lemma}

By \eqref{eqn55} we mean the following.  The product (on the left) of any
operator from the collection $(g\SC)\cap\BC$ with another
operator from the collection $(h\SC)\cap\BC$ belongs to the collection
$(gh\SC)\cap\BC$ (on the right), and every operator in $(gh\SC)\cap\BC$ can be written 
as such a product in precisely $M$ different ways.

We are now in a position to state and prove our central result:

\subsection{Isomorphism theorem}
\label{sbsct5B}

\begin{theorem}[Isomorphism]
\label{thm3}
  Let $C$ be an additive graph code with information group $\GC$, $P$ the projector onto the coding space $\HC_C$ and $\pt$ be some subset of the carrier qudits. Then the collection $\GC^\pt$ of members of $\GC$ with nonzero partial trace down to $\pt$, \eqref{eqn52}, is a subgroup of the information group $\GC$, and the mapping $\hat g\rightarrow g_{\pt}$ in \eqref{eqn51} carries $\GC^{\pt}$ to an
isomorphic group $\GC_{\pt}$ of nonzero operators on $\HC_\pt$.  Furthermore,

i) If $\hat g$ and $\hat h$ are any two elements of $\GC^\pt$, then
\begin{equation}
\label{eqn56} 
  \Tr_{\ptc}[\hat g \hat h] = \Tr_{\ptc}[\hat g]\;\Tr_{\ptc}[\hat h]/N \quad \text{or}\quad (gh)_\pt = g_\pt h_\pt 
\end{equation}

ii) If $\hat g\neq\hat h$ are distinct elements of $\GC^\pt$, 
$g_{\pt}\neq h_{\pt}$ are distinct elements of $\GC_\pt$. 

iii) The identity element
\begin{equation}
\label{eqn57}
P_{\pt}:=\Tr_{\ptc}[P]/N,
\end{equation}
of $\GC_{\pt}$ is a projector onto a subspace of $\HC_\pt$
(possibly the whole space) with rank equal to $\Tr[P]/N = K/N$. 

The normalization constant $N$ is given as
\begin{equation}
\label{eqn58}
 N := |\SC\cap\BC| \cdot D^{|\ptc|}/|\SC|
\end{equation}
where $\BC$ are the operators based in $\pt$.
\end{theorem}

\begin{proof}
The proof is a consequence of Lemma~\ref{thm2} and the following observations.
The trace $\Tr_{\ptc}[\hat g]$ in \eqref{eqn51} is, apart from a constant,
the trace of $\Sigma[(g\SC)\cap \BC]$, and is zero if $(g\SC)\cap \BC$ is
empty.  If the collection $(g\SC)\cap \BC$ is not empty, then by
Lemma~\ref{thm1} it consists of a collection of linearly independent
operators, and the trace of its sum cannot vanish. Thus there is a one-to-one,
see part (iii) of Lemma~\ref{thm2}, correspondence between nonempty sets of
the form $(g\SC)\cap\BC$ and the elements $\hat g$ in $\GC^\pt$. Then (i) and
(iv) of Lemma~\ref{thm2} imply both that $\GC^\pt$ is a group, and also that
the map from $\GC^\pt$ to $\GC_\pt$ is a homomorphism, whereas (ii) shows
that this is actually an isomorphism: $g_\pt=h_\pt$ is only possible when
$g\SC=h\SC$. That $N$ in \eqref{eqn58} is the correct normalization follows
from \eqref{eqn54}, \eqref{eqn55}, and \eqref{eqn48}.
\end{proof}

A significant consequence of Theorem~\ref{thm3} is the following result on the
presence and absence of information in the subset $B$, using
the terminology of Sec.~\ref{sct2}:

\begin{theorem} 
\label{thm4} 
Let $C$ be an additive graph code on $n$ carrier qudits, with information
group $\GC$. Let $\pt$ be a subset of the carrier qudits, $\GC^{\pt}$ the
corresponding subset information group, and $\JC(\hat g)$ the type of
information corresponding to $\hat g$ (as defined in Sec.~\ref{sct2}).
 Then
\begin{itemize}
 \item[i)] The $\JC(\hat g)$ type of information is perfectly
   present in $\pt$ if and only if $\hat g \in \GC^{\pt}$.
\item[ii)] The $\JC(\hat g)$ type of information is
absent from $\pt$ if and only if $\hat g^k \notin \GC^{\pt}$ for all integers
$k$ between $1$ and $D-1$.
 \item[iii)] All information is perfectly present in $\pt$ if and only if
   $\GC^{\pt}=\GC$.
 \item[iv)] All information is absent from $\pt$ if and only if $\GC^{\pt}$
   consists entirely of scalar multiples of the identity element $P$ of $\GC$.

\end{itemize}
\end{theorem}

The proof of the theorem can be found in App.~\ref{apdxB}. Statement (iii) is useful because
the check of whether there is a perfect quantum channel from the input
to $\pt$ involves a finite group $\GC$; one does not have to consider 
all normal operators of the form \eqref{eqn2}. 
Statement (ii) deserves further comment.  If $D$ is prime then the order of
any element of the Pauli group (apart from the identity) is $D$, see the
remark following \eqref{eqn9}. The same is true of elements of the quotient
group $\GCbar$, \eqref{eqn41}, and thus of members $\hat g$ of the isomorphic
group $\GC$.  Consequently, for any $k$ in the interval $1 < k < D$, there is
some $m$ such that $1=km \mod D$, which means $\hat g = (\hat g^k)^m$.  And
since $\GC^\pt$ is a group, $\hat g^k \in\GC^\pt$ implies $\hat
g\in\GC^\pt$. Thus when $D$ is prime, $\hat g\notin \GC^\pt$ is equivalent to
$\hat g^k \notin \GC^{\pt}$ for all integers $k$ between $1$ and $D-1$, and
the latter can be replaced by the former in statement (ii).
However, when $D$ is composite it is quite possible
to have $\Tr_{\ptc}[\hat g] = 0$ but $\Tr_{\ptc}[\hat g^{k'}] \neq 0$ for some
$k'$ larger than 1 and less than $D$; see the example below. In this situation
we can still say that $\JC(\hat g^{k'})$ is perfectly present, but it is not
true that $\JC(\hat g)$ is absent. One can regard the type $\JC(\hat g)$ as a
\emph{refinement} of $\JC(\hat g^{k'})$, and as explained in Sec.~\ref{sct2},
although the coarse-grained $\JC(\hat g^{k'})$ information is perfectly
present in $\pt$, the additional information associated with the refinement is
not.

As an example, suppose $\hat g$ has a spectral decomposition
\begin{equation}
\label{eqn59}
\hat g=J_0+\ii J_1-J_2-\ii J_3,
\end{equation}
with the $J_j$ orthogonal projectors such that
\begin{align}
\label{eqn60}
\Tr_{\ptc}[J_0]=\Tr_{\ptc}[J_2]\neq\Tr_{\ptc}[J_1]=\Tr_{\ptc}[J_3].
\end{align}
Then $\Tr_{\ptc}[\hat g]=0$, whereas 
\begin{equation}
\label{eqn61}
\hat g^2=(J_0+J_2)-(J_1+J_3),
\end{equation}
and thus $\Tr_{\ptc}[\hat g^2]\neq 0$.  Thus $\hat g^2$ is an element of
$\GC^{\pt}$, whereas $\hat g$ is not, and so the coarse grained $\JC(\hat g^2)$
information corresponding to the decomposition on the right side of
\eqref{eqn61} is present in $\pt$, while the further refinement corresponding
to the right side of \eqref{eqn59} is not.  Precisely this structure is
produced by a graph code on two carriers of dimension $D=4$, with graph state
$\ket{G}=\ket{++}$, coding group $\CC=\langle Z_1Z_2\rangle$, information
group $\GC=\langle X_1P , Z_1Z_2P \rangle$, coding space projector
\begin{equation}
\label{eqn62}
 P = (I + X_1 X_2^{3} + X_1^2 X_2^2 +X_1^3 X_2)/4,
\end{equation}
and 
\begin{equation}
\label{eqn63}
\hat g=X_1P=\dyad{\bar 0\bar0}{\bar0\bar0}+
\ii\dyad{\bar 1\bar 2}{\bar 1 \bar 2}-\dyad{\bar 2\bar0}{\bar 2\bar0}
-\ii\dyad{\bar 3\bar 2}{\bar 3 \bar 2},
\end{equation}
where $\ket{\bar j}=Z^j\ket{+}$ are the eigenvectors of the $X$ operator.

\subsection{Information flow}
\label{sbsct5C}

At this point let us summarize how we think about information ``flowing'' from
the input via the encoding operation into a subset $\pt$ of the code
carriers. At the input the information is represented by the quotient group
$\GCbar_0=\WC_0/\SC_0$, see \eqref{eqn41}, or more concretely by the
isomorphic group $\GC_0$ of operators generated by the cosets, as in
\eqref{eqn48}.  The encoding operation $UW$, see \eqref{eqn36} and
\eqref{eqn38}, maps $\GCbar_0$ to the analogous $\GCbar=\WC/\SC$ associated
with the code $C$, and likewise $\GC_0$ to the group of operators $\GC$ acting
on the coding space $\HC_C$.  Tracing away the complement $\ptc$ of $\pt$ maps
some of the $\hat g$ operators of $\GC$ to zero, and the remainder form the
subset information group $\GC^{\pt}$. Applying the inverse $UW$ map to
$\GC^{\pt}$ gives $\GC^{\pt}_0$, a subgroup of $\GC_0$ that tells us what
types of information at the input (i.e. before the encoding) are available in
the subset of carriers $\pt$.  This is illustrated by various examples in the
next section.

\section{Examples}
\label{sct6}

\subsection{General principles}
\label{sbsct6A}

In this section we apply the principles developed earlier in the paper to some
simple $[[n,k,\delta]]_D$ additive graph codes, where $n$ is the number of
qudit carriers, each of dimension $D$, the dimension of the coding space
$\HC_C$ is $K=D^k$, and $\delta$ is the distance of the code; see Chapter 10 of \cite{NielsenChuang:QuantumComputation} for a
definition of $\delta$.  We shall be interested in the subset information
group $\GC^\pt$, \eqref{eqn52}, that represents the information about the
input that is present in the subset $\pt$ of carriers.  Rather than discussing
$\GC^\pt$ or its traced down counterpart $\GC_\pt$, it will often be simpler
to use $\GC_0^\pt$, the subset information group referred back to the channel
input, see Sec.~\ref{sbsct5C} above, and in this case we add an initial
subscript $0$ to operators: $X_{01}$ means the $X$ operator on the first qudit
of the input.  Since all three groups are isomorphic to
one another, the choice of which to use in any discussion is a matter of
convenience. (In the examples below for the sake of brevity we sometimes 
omit a term $\mathrm{e}^{\ii \phi}I$ from the list of generators of $\GC^{\pt}_0$.)

Before going further it is helpful to list some general principles of quantum
information that apply to all codes, and which can simplify the analysis of
particular examples, or give an intuitive explanation of why they work.  In
the following statements ``information'' always means information about the
input which has been encoded in the coding space through some isometry.

1. If all information is perfectly present in $\pt$,
then all information is absent from $\ptc$.

2. If all information is absent from $\ptc$ then all information is perfectly
present in $\pt$.

3. If the information about some orthonormal basis (i.e., the type
corresponding to this decomposition of the identity) is perfectly present in
$\pt$, then the information about a mutually-unbiased basis is absent from
$\ptc$.

4. If two types of information that are ``sufficiently incompatible'' are both
perfectly present in $\pt$, then all information is perfectly present in
$\pt$.  In particular this is so when the two types are associated with
mutually unbiased bases.

5. For a code of distance $\delta$ all information is absent from any $\pt$ if
$|\pt|<\delta$, and all information is perfectly present in $\pt$ if $|\pt| >
n-\delta$.

Items 1, 2, 3 and 4 correspond to the No Splitting, Somewhere, Exclusion
and Presence theorems of \cite{PhysRevA.76.062320}, which also gives weaker
conditions for ``sufficiently incompatible.''  The essential idea behind 5 is
found in Sec.~III~A of \cite{PhysRevA.56.33}
\footnote{It is shown in \cite{PhysRevA.56.33} that if noise only affects a certain 
subset $\ptc$ of the carriers with $|\ptc|<\delta$, then the errors can be corrected
using the complementary set $\pt$. In our notation this is equivalent to saying that all the information is in $\pt$.
}.

\subsection{One encoded qudit \label{sbsct6B}}

\begin{figure}
\includegraphics{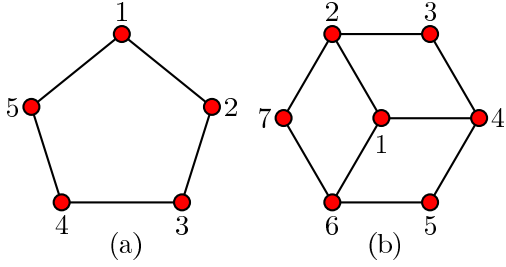}
\caption{(a) The graph state for the $[[5,1,3]]_D$ code; (b) The graph state
  for Steane $[[7,1,3]]_2$ code}
\label{five_steane}
\end{figure}

It was shown in \cite{IEEETransInfTheory.45.1827} that a $[[5,1,3]]_D$
code exists for all $D$. Here we consider the graph version
\cite{PhysRevA.65.012308} where the coding group is
\begin{equation}
\label{eqn64}
\CC=\langle Z_1 Z_2 Z_3 Z_4 Z_5\rangle
\end{equation}
and the graph state is shown in Fig.~\ref{five_steane}(a). Our formalism shows
that, whatever the value of $D$, there are only two possibilities.  When $|B|$
is 1 or 2 $\GC^{\pt}$ is the just the group identity, the projector $P$ on the
coding space, so all information is absent whereas if $|B|$ is 3, 4 or
(obviously) 5, $\GC^{\pt} = \GC$, so the subsystem $B$ is the output of a
perfect quantum channel.  To be sure, these results also follow from principle
5 in the above list, given that $\delta=3$ for this code.

The Steane $[[7,1,3]]_2$ code, a graphical version of which
\cite{quantph.0709.1780} has a coding group
\begin{equation}
\label{eqn65}
\CC=\langle Z_3 Z_5 Z_7\rangle
\end{equation}
for the graph state shown in Fig.~\ref{five_steane} (b), is more interesting
in that while principles 5 ensures that all $|\pt|\leq 2 = \delta -1$ subsets
of carriers contain zero information and all $|\pt|\geq 5 = n-\delta+1$
subsets contain all the information, one qubit, it leaves open the question of
what happens when $|\pt|=3$ or 4. We find that all information is perfectly
present when $\pt$ is $\{1,2,5\}$, $\{1,3,6\}$, $\{1,4,7\}$, $\{2,3,4\}$,
$\{2,6,7\}$, $\{4,5,6\}$, or $\{3,5,7\}$---representing three different
symmetries in terms of the graph in the figure---and absent for all other
cases of $|B|=3$.  Therefore all information is absent from the $|\pt|=4$ subsets which
are complements of the seven just listed, and perfectly present in all others
of size $|\pt|=4$.  So far as we know, generalizations of this code to $D>2$
have not been studied.

\begin{figure}
\includegraphics{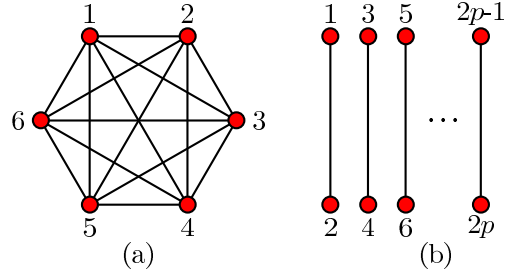}
\caption{(a) Complete graph (on $6$ qudits); (b) Bar graph with $n=2p$ carriers and $p$ bars}
\label{bar_complete}
\end{figure}

A simple code in which a specific type of information is singled out is
$[[n, 1, 1]]_D$ generated by
\begin{equation}
\label{eqn66}
\CC=\langle Z_1Z_2 \dotsm Z_n\rangle
\end{equation}
on the \emph{complete graph}, illustrated in Fig.~\ref{bar_complete}(b) for
$n=6$.  Whereas all information is (of course) present when $|\pt|=n$, it
turns out that for any subset $B$ with $1\leq |\pt| < n$ one has 
$\GC^{\pt}_{0}= \langle X_{01} Z_{01} \rangle$, i.e., the abelian group consisting
of all powers of the operator $X_1 Z_1$ on the input qudit. 
Thus the information is ``classical,'' corresponding to that
decomposition of the input identity that diagonalizes $X_1 Z_1$.
The intuitive explanation for this situation is that this $X_1 Z_1$ type
of information is separately copied as an ideal classical channel, see \eqref{eqn6},
 to each of the carrier qudits, and as a
consequence other mutually unbiased types of information are ruled out by
principle 3. This, of course, is typical of ``classical'' information, which
can always be copied. 

A more interesting example in which distinct types of information come into
play is the bar graph, Fig.~\ref{bar_complete} (a), in which $n$ qudits are
divided up into $p=n/2$ pairs or ``bars,'' and the code is generated by 
\begin{equation}
\label{eqn67}
\CC=\langle Z_1Z_2\dotsm Z_n\rangle.
\end{equation}
Let us say that a subset of carriers $\pt$ has property I if the corresponding
subgraph contains at least one of bars, and property II if it contains at
least one qudit from each of the bars.  Then:

(i) If $\pt$ has property I but not II, $\GC^{\pt}_{0} = \langle
X_{01} \rangle$, an abelian group.

(ii) If $\pt$ has property II but not I, $\GC^{\pt}_{0} = \langle
X_{01}^p Z_{01} \rangle$, another abelian group

(iii) If $\pt$ has both property I and property II, all information (1 qudit)
is perfectly present. 

(iv) When $\pt$ has neither property I nor II, all information is absent.

While both (i) and (ii) are ``classical'' in an appropriate sense and indeed represent an ideal classical channel, the two
abelian groups do not commute with each other, so the two types of information
are incompatible, and it is helpful to distinguish them. Case (iii)
illustrates principle 4, since $X_{01}$ and $X_{01}^p Z_{01}$ (whatever the value
of $p$) correspond to mutually unbiased bases. In case (iv) the complement
$\ptc$ of $\pt$ possesses both properties I and II, and therefore contains all
the information, so its absence from $\pt$ is an illustration of principle 1.

\subsection{Two encoded qudits\label{sbsct6C}}

\begin{figure}
\includegraphics{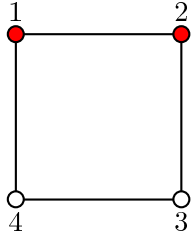}
\caption{The graph state  of the $[[4,2,2]]_D$ code}
\label{K4code}
\end{figure}

Consider a $[[4,2,2]]_D$ code based on the graph state shown in Fig.~\ref{K4code} whose
coding group 
\begin{equation}
\label{eqn68}
\CC=\langle Z_1 Z_2, Z_3 Z_4 \rangle,
\end{equation}
employs two generators of order $D$, and thus encodes two qudits.  Note that
while the graph state has the symmetry of a square the coding group has a
lower symmetry corresponding to the different types of nodes employed in the
figure.

Let us begin with the qubit case $D=2$.  Our analysis shows that
when $|\pt|=1$ all information is absent, and thus for $|\pt|\geq 3$ all
information is present, consistent with the fact that this code has
$\delta=2$ \cite{PhysRevA.78.042303},  see principle 5. Thus the
interesting cases are those in which $|\pt|=|\ptc|=2$, for which one finds: 
\begin{align}
\label{eqn69}
 \pt=\{1,3\}, \ptc=\{2,4\}:&\quad  
 \GC^{\pt}_{0}=\GC^{\ptc}_{0} = \langle X_{01} Z_{01} Z_{02}, X_{01} X_{02}  \rangle;\\ 
\label{eqn70}
\pt=\{1,4\}, \ptc=\{2,3\}:&\quad  
 \GC^{\pt}_{0}=\GC^{\ptc}_{0} = \langle X_{01} Z_{01}, X_{02} Z_{02} \rangle;\\ 
\label{eqn71}
 \pt=\{1,2\}, \ptc=\{3,4\}:&\quad  
 \GC^{\pt}_{0}=\GC^{\ptc}_{0} = \langle X_{01} Z_{01}, X_{02} Z_{02} \rangle.
\end{align}
In each case the generators commute and thus the subgroup $\GC^{\pt}_{0}$ is
abelian. Hence the information is ``classical'', and the same type is present
both in $\pt$ and $\ptc$, not unlike the situation for the complete graph
considered earlier. However, the three subgroups do not commute with each
other, so the corresponding types of information are mutually incompatible, a
situation similar to what we found for the bar graph.

For $D>2$ it is again the case that all information is absent when $|\pt|=1$
completely present for $|\pt|\geq 3$.  And \eqref{eqn69} and \eqref{eqn70}
remain correct (with each generator of order $D$), and these subgroups are
again abelian.  However, when $\pt=\{1,2\}$ and $\ptc=\{3,4\}$, \eqref{eqn71}
must be replaced with
\begin{equation}
\label{eqn72}
\GC^{\pt}_{0} = \langle Z_{01}^{} X_{02}^{2}, Z_{02}^{} \rangle,\quad 
\GC^{\ptc}_{0} = \langle Z_{01}^{}, X_{01}^2 Z_{02}^{} \rangle.
\end{equation}
In each case the two generators do not commute with each other, so neither
subgroup is abelian.  However, all elements of $\GC^{\pt}_{0}$ commute with
all elements of $\GC^{\ptc}_{0}$. Also, the two subgroups are isomorphic
(interchange subscripts 1 and 2).

For \emph{odd} $D \geq 3$ one can use for $\GC^{\pt}_{0}$ an alternative pair
of generators
\begin{equation}
\label{eqn73}
\GC^{\pt}_{0} = \langle Z_{01}^m X_{02}^{}, Z_{02}^{} \rangle,\quad 
m:=(D+1)/2,
\end{equation}
whose order is $D$ and whose commutator is
\begin{equation}
\label{eqn74}
\bigl( Z_{01}^m X_{02}^{}\bigr) Z_{02} = \omega Z_{02} \bigl( Z_{01}^m X_{02}^{}\bigr).
\end{equation}
This means---see \eqref{eqn8}---that $\GC^{\pt}_{0}$, and thus also the
(isomorphic) $\GC^{\ptc}_{0}$, is isomorphic to the
Pauli group of a single qudit. Since  $\GC^{\pt}_{0}$ and
$\GC^{\ptc}_{0}$ commute with each other, it is natural to think of the pair
as associated with the tensor product of two qudits with the same $D$.  That
this is correct can be confirmed by explicitly constructing
a ``pre-encoding'' circuit embodying the unitary
\begin{equation}
\label{eqn75}
(\mathrm{F}_1\otimes \mathrm{F}_2)^\dag \mathrm{CP}^{-m}_{12}
(\mathrm{F}_1\otimes \mathrm{F}_2),
\end{equation} 
expressed in terms of the Fourier and CP gates defined in Sec.~\ref{sbsct3B},
that carries the Pauli groups on ``pre-input" qudits 1 and 2 onto $\GC^{\pt}_{0}$ and
$\GC^{\ptc}_{0}$, respectively. 

Things become more complicated for \emph{even} $D\geq 4$, where $\GC^{\pt}_{0}$ (and also $\GC^{\ptc}_{0}$) are no longer isomorphic to the Pauli group of a single qudit.

\section{Conclusion}
\label{sct7}

We have shown that for additive graph codes with a set of $n$ carrier qudits,
each of the same dimension $D$, where $D$ is any integer greater than 1, it is
possible to give a precise characterization of the information from the
coding space that is present in an arbitrary subset $\pt$ of the carriers.
This information corresponds to a subgroup $\GC^\pt$ of a group $\GC$, the
information group of operators on the coding space, that spans the coding
space and provides a useful representation of the information that it
contains.  We discuss how what we call a trivial code, essentially a tensor
product of qudits of (possibly) different dimensions, can be encoded into the
coding space in a manner which gives one a clear intuitive interpretation of
$\GC$.  The subgroup $\GC^\pt$ is then simply the subset of operators in $\GC$
whose trace down to $\pt$ is nonzero, and the traced-down operators when
suitably normalized form a group $\GC_\pt$ that is isomorphic to $\GC^\pt$.
The information present in those operators in $\GC$ that are not in $\GC^\pt$
disappears so far as the subsystem $\pt$ is concerned, as their partial traces
are zero.  This is the central result of our paper and is illustrated by a
number of simple examples in Sec.~\ref{sct6}. We also provide in
App.~\ref{apdxC} a relatively simple algorithm for finding the elements of
$\GC^B$.

These results can be extended to arbitrary qudit stabilizer codes even if they
are not graph codes, by employing appropriate stabilizer and information
groups, as in Sec.~\ref{sct4}. Here, however, the concept of a trivial code,
and thus our perspective on the encoding step, may not apply.  The extension
of these ideas, assuming it is even possible, to more general codes, such as
nonadditive graph codes, remains an open question.

As shown in App.~\ref{apdxD} our formalism can be fitted within the general
framework of invariant algebras as discussed in \cite{PhysRevLett.98.100502,
  PhysRevA.76.042303,arxiv0907.4207,PhysRevLett.100.030501}.  The overall
conceptual framework we use is somewhat different from that found in these
references in that we directly address the question of what information is
present in the subsystem of interest, rather than asking whether there exists
some recovery operation (the $\RC$ in App.~\ref{apdxD}) that will map an
algebra of operators back onto its original space. Thus in our work the
operator groups $\GC^\pt$ on the coding space and $\GC_\pt$ on the subsystem
are isomorphic but not identical.  Hence, even though there is, obviously, a
close connection between our ``group approach'' and the ``algebraic
approach,'' the algebra of interest being generated from the group of
operators, further relationships remain to be explored.  The fact that the
arguments in App.~\ref{apdxD} are not altogether straightforward suggests that
the use of groups in cases where this is possible may provide a useful
supplement, both mathematically and intuitively, to other algebraic ideas.  In
particular the additional structure present in an additive graph code allows
one to determine $\GC^\pt$ in $\OC(n^{\theta}+K^2n^2)$, App.~\ref{apdxC}, as
against $\OC(K^6)$ for the algorithm presented in
\cite{PhysRevLett.100.030501} for a preserved matrix algebra, where $K$ is the
dimension of the input and output Hilbert space.

\section*{Acknowledgments}
We thank Chang-You Lin for his contributions and Li Yu for useful discussions. The research described here received support from the National Science Foundation through Grants No. PHY-0456951 and PHY-0757251.


\appendix
\section{Proof of Lemmas~\ref{thm1} and \ref{thm2} }
\label{apdxA}

Proof of Lemma~\ref{thm1}

The operators in $p\RC$ are linearly independent when those in $\RC$ are
linearly independent, since $p$ is unitary and thus invertible.  This
establishes (i).  For (ii), consider the case where $q$ is the identity $I$.
As the collection $\RC$ is linearly independent, there is at most one $r\in
\RC$ such that $pr$ is a multiple of the identity.  If such an $r$ exists, $p$
is of the form $\mathrm{e}^{\ii\phi} r^{-1}$, and since $\RC$ is a group,
$p\RC=\mathrm{e}^{\ii\phi}r^{-1}\RC=\mathrm{e}^{\ii\phi}\RC$, we have situation $(\alpha)$, with the collection
$p\RC\cup\RC$ linearly dependent.  Next assume the collection
$p\RC\cup\RC$ is linearly dependent, which means there are complex numbers
$\{a_r\}$ and $\{b_r\}$, not all zero, such that
\begin{equation}
\label{eqn76} 
\sum_{r\in\RC} \left[ a_r r + b_r pr\right] = 0. 
\end{equation}
This is not possible if all the $a_r$ are zero, since this would mean $p\sum_r
b_r r=0$, thus $\sum_r b_r r=0$ implying $b_r=0$ for every $r$, since the
$\RC$ collection is by assumption linearly independent.  Thus at least one
$a_r$, say $a_s$ is not zero. Multiply both sides of \eqref{eqn76} by $s^{-1}$
on the right and take the trace:
\begin{equation}
\label{eqn77}
a_s\Tr[I] + \sum_{r\in\RC} b_r\Tr[prs^{-1}] = 0, 
\end{equation}
implying there is at least one $r$ for which $\Tr[prs^{-1}]\neq 0$.  But then
$p$ is of the form $\mathrm{e}^{\ii\phi} sr^{-1}= \mathrm{e}^{\ii\phi}\bar r^{-1}$ for 
$\bar r = rs^{-1}\in\RC$, so we are back to situation $(\alpha)$. Hence the alternative
to $(\alpha)$ is $(\beta)$: the operators in $p\RC\cup\RC$ are linearly
independent.  Finally, if $q$ is not the identity $I$, simply apply the
preceding argument with $\bar p = q^{-1}p$ in place of $p$.
\medskip

Proof of Lemma~\ref{thm2}

Statement (i) is a consequence of the fact that if an invertible operator is in
$\BC$, so is its inverse, and since
$\SC$ is a group, $g\SC$ consists of the inverses of the elements in
$g^{-1}\SC$.

Statements (ii) and (iv) follow from a close examination of \eqref{eqn55}.
Assume both sets on the left side are nonempty. If $gs_1$ 
and  $hs_2$ are both in $\BC$, so is their product $gs_1hs_2= ghs_1s_2$, where
we use the fact that $g$ and $h$ commute with every element of $\SC$. 
If, on the other hand, $(gh\SC)\cap\BC$ and $(g\SC)\cap\BC$ are nonempty,
any element, say $ghs_1$, in the former can be written using a specific
element, say $g\bar s$, in the latter, as
\begin{equation}
\label{eqn78}
 ghs_1 = (g\bar s)(h s_2) 
\end{equation}
where $s_2 = s_1 \bar s^{-1}$ is uniquely determined by this equation, and the
fact that both $ghs_1$ and $g\bar s$ are (by assumption) in $\BC$ means the
same is true of $h s_2$.  Thus not only can every element of $(gh\SC)\cap\BC$
be written as a product of elements of $(g\SC)\cap\BC$, but there is a one-to-one
correspondence between $(gh\SC)\cap\BC$ and $(g\SC)\cap\BC$, which must
therefore be of equal size. A similar argument shows that $(gh\SC)\cap\BC$
and  $(h\SC)\cap\BC$ are of the same size. This establishes both (ii)
and (iv).

As for (iii), use the fact that the cosets $g\SC$ and $h\SC$ are either
identical or have no elements in common, so the same is true of their
intersections with $\BC$.  If $g\SC$ and $h\SC$ have no elements in common,
Lemma~\ref{thm1} with $\RC=\SC$ tells us that either $g\SC= \mathrm{e}^{\ii\phi} (h\SC)$
for some nonzero $\phi$, in which case 
$\Sigma[(g\SC)\cap\BC] = \mathrm{e}^{\ii\phi}\Sigma[(h\SC)\cap\BC]$
is distinct from $\Sigma[(h\SC)\cap\BC]$, or else the collection $(g\SC)\cup
(h\SC)$ is linearly independent, which means that its intersection with $\BC$
shares this property and the operators $\Sigma[(g\SC)\cap\BC]$ and
$\Sigma[(h\SC)\cap\BC]$ are linearly independent.

\section{Proof of  Theorem~\ref{thm4} \label{apdxB}}

The proof of Theorem~\ref{thm4} makes use of the following:
\begin{lemma}
\label{thm5}

Let $\hat g=P\hat gP$ be an information operator in $\GC$ with spectral decomposition
\begin{equation}
\label{eqn79} 
\hat g  = \sum_{j=0}^{m-1} \lambda_j J_j, 
\end{equation}
where the mutually orthogonal projectors $J_j$ sum to $P$.  Then each
projector $J_j$ can be written as a polynomial in $\hat g$ with $\hat g^0=P$:
\begin{equation}
\label{eqn80} 
J_j = \sum_{k=0}^{m-1} \alpha_{jk} \hat g^k.
\end{equation}
\end{lemma}

\begin{proof}
The proof consists in noting that
\begin{equation}
\label{eqn81} 
\hat g^k = \sum_{j=0}^{m-1} \lambda_j^k J_j=\sum_{j=0}^{m-1}\beta_{kj}J_j,
\end{equation}
is a linear equation in the $J_j$ with $\beta_{kj} = \lambda_j^k$ an $m\times
m$ Vandermonde matrix whose determinant is $\prod_{j > k}(\mu_j-\mu_k)$ (see
p.~29 of \cite{HornJohnson:MatrixAnalysis}).  As the $\mu_j$ are distinct the
matrix $\beta_{kj}$ has an inverse $\alpha_{jk}$.
\end{proof}

To prove (i) of Theorem~\ref{thm4}, first assume that $\hat g$ is in
$\GC^\pt$.  Since $\GC^{\pt}$ is a group with identity $P$, this means that all
powers of $\hat g$, including $\hat g^0=P$, are also in $\GC^{\pt}$.
Consequently, the projectors entering the spectral decomposition \eqref{eqn79}
of $\hat g$ satisfy
\begin{equation}
\label{eqn82}
N^{-1}\Tr_{\ptc}[J_j] \; \Tr_{\ptc}[J_k]
= \Tr_{\ptc}[J_jJ_k] 
= \delta_{jk}\Tr_{\ptc}[J_j],
\end{equation}
with the first equality obtained by expanding $J_j$ and $J_k$ in powers of
$\hat g$, \eqref{eqn80}, and using \eqref{eqn56} along with the linearity of
the  partial trace.  This orthogonality of the partial traces of different projectors,
see \eqref{eqn3}, implies that the $\JC(\hat g)$ type of information is
perfectly present in $\pt$. Conversely, if the $\JC(\hat g)$ type of
information is perfectly present in $\pt$ then the partial traces down to
$\pt$ of the different $J_j$, which cannot be zero, are mutually orthogonal
and thus linearly independent.  Therefore by \eqref{eqn79}, $\Tr_{\ptc}[\hat
g]$ cannot be zero, and $\hat g$ is in $\GC^\pt$.

The prove (ii) note that $\hat g^k$ absent from $\GC^{\pt}$ for $1\leq k < D$ 
means that 
$\Tr_{\ptc}[\hat g^k]=0$ for these values of $k$, and thus by taking the
partial trace of both sides of  \eqref{eqn80} and using \eqref{eqn57},  
\begin{equation}
\label{eqn83} 
\Tr_{\ptc}[J_j] = N\alpha_{j0} P_{\pt}.
\end{equation}
Since these partial traces are identical up to a multiplicative constant 
there is no information of the $\JC(\hat g)$ type in $\pt$.  For the converse,
if there is no $\JC(\hat g)$ information in $\pt$ then there is also no
$\JC(\hat g^2)$, $\JC(\hat g^3)$, etc. information in $\pt$, since the
projectors which arise in the spectral decomposition of $\hat g^k$ are already
in the spectral decomposition of $\hat g$, see \eqref{eqn81}.
Consequently, by (i), these $\hat g^k$ must be absent from $\GC^\pt$.  

To prove (iii), note that if all information is perfectly present in $\pt$
this means that for every $\hat g\in\GC$ the $\JC(\hat g)$ information is
present in $\pt$, and therefore, by (i), $\hat g\in\GC^\pt$, so $\GC =
\GC^\pt$. For the converse, let $Q_1$ and $Q_2$ be two orthogonal but
otherwise arbitrary projection operators on subspaces of the coding space
$\HC_C$.  Because the elements of the information group $\GC$ form a basis for
the set of linear operators on $\HC_C$, see comments at the end of
Sec.~\ref{sbsct4C}, $Q_1$ and $Q_2$ can both be written as sums of elements
$\hat g$ in $\GC$, and the same argument that was employed in \eqref{eqn82}
shows that the orthogonality of $Q_1$ and $Q_2$ implies the orthogonality of
$\Tr_{\ptc}[Q_1]$ and $\Tr_{\ptc}[Q_2]$.

To prove (iv), note that if $\GC^{\pt}$ consists entirely of scalar
multiples of $P$, the partial trace down to
$\pt$ of any projector $Q$ on a subspace of $\HC_C$, since it can be written
as a linear combination of the partial traces of the $\hat g$ in $\GC$, most
of which vanish, will be some multiple of $P_{\pt}$, and thus all information is
absent from $\pt$.  Conversely, if $\GC^{\pt}$ contains a $\hat g$ which is
not proportional to $P$ the corresponding $\JC(\hat g)$ type of information
will be present in $\pt$ by (i), so it is not true that all information is
absent from $\pt$, a contradiction.

\section{Algorithm for Finding $\GC^{\pt}$\label{apdxC}}

Here we present an algorithm for determining the subset information group
$\GC^{\pt}$ by finding the elements $\hat g$ of $\GC$ whose partial trace down
to $\pt$ is nonzero.  If two or more elements differ only by a phase it is
obviously only necessary to check one of them.  For what follows it is helpful
to adopt the abbreviation
\begin{equation}
\label{eqn84}
E^{(\vect{x}|\vect{z})} := X^{\vect{x}} Z^{\vect{z}}
\end{equation}
with $(\vect{x}|\vect{z})$ an $n$-tuple row
vector pair, and thus a $2n$-tuple of integers between $0$ and $D-1$. 
Arithmetic operations in the following analysis are assumed to be mod $D$.

First consider the trivial code on the trivial graph, Sec.~\ref{sbsct4B}, with
information group $\GC_0^\pt$ consisting of elements of the form $\hat g_0=
g_0 P_0$, see \eqref{eqn48}, with $g_0 = E^{(\vect{x}_0|\vect{z}_0)}$ some 
element of $\WC_0=\langle \SC_0^G, \CC_0\rangle$, and
\begin{equation}
P_0 = |\SC_0|^{-1}\sum_{\vect{x}\in\XZ} X^{\vect{x}},
\label{eqn85} 
\end{equation}
where $\XC_0$ denotes the collection of $n$-tuples that enter the stabilizer
$\SC_0$, \eqref{eqn37}.
 By choosing $\vect{x}_0$ and $\vect{z}_0$ to be of the form
\begin{align}
\label{eqn86}
\vect{x}_0 &= (\xi_1,\xi_2,\ldots \xi_k,0,0,\ldots 0),\notag\\
\vect{z}_0 &= (\zeta_1 m_1,\zeta_2 m_2,\ldots \zeta_k m_k,0,0,\ldots 0),
\end{align}
using integers in the range
\begin{equation}
\label{eqn87}
 0\leq \xi_j \leq (d_j-1),\quad  0\leq \zeta_j \leq (d_j-1), 
\end{equation}
we obtain a single representative $g_0 =E^{(\vect{x}_0|\vect{z}_0)}$ for each
coset $g_0\SC_0$ in $\WC/\SC_0$.  The corresponding information operator,
which depends only on the coset, is 
\begin{equation}
\label{eqn88} 
\hat g_0 =  E^{(\vect{x}_0|\vect{z}_0)} P_0 = 
 |\SC_0|^{-1} \sum_{\vect{x}\in\XZ} \omega^{-\vect{z}_0\vect{x}} 
 E^{(\vect{x}+\vect{x}_0|\vect{z}_0)},
\end{equation}
where the addition of $\vect{x}$ and $\vect{x}_0$ is component-wise mod $D$,
and $\vect{z}_0\vect{x}$ denotes the scalar product of $\vect{z}_0$ and
$\vect{x}$ mod $D$ (multiply corresponding components and take the sum mod
$D$).

Elements of the information group $\GC^\pt$ of the nontrivial code of interest
to us are then of the form
\begin{align}
\label{eqn89}
\hat g &= (UW)\hat g_0(UW)^{\dagger}\notag\\
 &=  
|\SC_0|^{-1} \sum_{\vect{x}\in\XZ}
\omega^{\nu(\vect{x},\vect{x}_0,\vect{z})-\vect{z}_0\vect{x}}
E^{(\vect{x}+\vect{x}_0|\vect{z}_0)\QZ},
\end{align}
where we use the fact that because the conjugating operator $UW$,
\eqref{eqn36}, is a Clifford operator there is a $2n\times 2n$ matrix $Q$ over
$\ZZ_D^{2n}$, representing a symplectic automorphism
\cite{PhysRevA.71.042315}, such that
\begin{equation}
\label{eqn90} 
(UW) E^{(\vect{x}|\vect{z})}(UW)^{\dagger} = 
 \omega^{\nu(\vect{x},\vect{z})} E^{(\vect{x}|\vect{z})\QZ}.
\end{equation}
with $(\vect{x}|\vect{z})Q$ the $2n$-tuple, interpreted as an $n$-tuple pair,
obtained by multiplying $(\vect{x}|\vect{z})$ on the right by $Q$, and
$\nu(\vect{x},\vect{z})$ an integer whose value does not concern us. The
explicit form of $Q$ can be worked out by means of the encoding procedure
presented in Sec.~\ref{sbsct4B}, using Tables~\ref{tbl1} and~\ref{tbl2}.

The operators appearing in the sum on the right side of \eqref{eqn89}
are linearly independent Pauli products, since $Q$ is nonsingular. The 
trace down to $\pt$ of such a product is nonzero if and only if its base is in 
$\pt$, and when  nonzero the result after the trace is essentially the same
operator: see \eqref{eqn53} and the associated discussion.  Consequently
$g_B=N^{-1} \Tr_{\ptc}[\hat g]$ is nonzero if and only if the trace down to
$\pt$ of at least one operator on the right side of \eqref{eqn89} is nonzero.
A useful test takes the form
\begin{equation}
\label{eqn91} 
\Tr_{\ptc}[ E^{(\vect{x}|\vect{z})} ]\neq 0
\Longleftrightarrow (\vect{x}|\vect{z})J  =\vect{0},
\end{equation}
where $\vect{0}$ is the zero row vector, and $J$ is a diagonal $2n\times 2n$
matrix with 1 at the diagonal positions $j$ and $2j$ whenever qudit $j$
belongs to $\ptc$, and 0 elsewhere.  Therefore the $\hat g$ associated with
$\vect{x}_0$ and $\vect{z}_0$ through \eqref{eqn88} and \eqref{eqn89} is a
member of $\GC^B$ if and only if there is at least one $\vect{x}\in\XC_0$
such that
\begin{equation}
\label{eqn92}
(\vect{x}+\vect{x}_0|\vect{z}_0)Q J 
= \vect{0} \text{\ \ or\ \ } (\vect{x}|\vect{0})Q J 
= -(\vect{x}_0|\vect{z}_0)Q J.
\end{equation}

The $\vect{x}$ that belong to $\XZ$ are characterized by the equation
\begin{equation}
\label{eqn93}
\vect{x} M = \vect{0},
\end{equation}
where $M$ is an $n\times k$ matrix that is everywhere 0 except for
$M_{jj}=m_j$ for $1\leq j\leq k$, using the $m_j$ that appear in
\eqref{eqn28}.  Consequently, instead of asking whether \eqref{eqn92} has a
solution $\vect{x}$ belonging to $\XZ$ one can just as well ask if there is
any solution to the pair \eqref{eqn92} and \eqref{eqn93}, or equivalently to
the equation
\begin{equation}
\label{eqn94}
\vect{x} T = \vect{u}_0 
\end{equation}
where $T$ is an $n\times (2n+k)$ matrix whose first $2n$ columns consist of
the top half of the matrix $QJ$, (upper $n$ elements of each column), and
whose last $k$ columns are the matrix $M$ in \eqref{eqn93}; while
$\vect{u}_0$ is a row vector whose first $2n$ elements are
$-(\vect{x}_0|\vect{z}_0)Q J$ and last $k$ elements are 0. Deciding if
\eqref{eqn94} has a solution $\vect{x}$ becomes straightforward once one has
transformed $T$ to Smith normal form, including determining the associated
invertible matrices, see \eqref{eqn32}.  As this needs to be done just once
for a given additive code and a given subset $\pt$, the complexity of the
algorithm for finding $\GC^\pt$ is $\OC(n^{\theta})$ for finding the Smith
form plus $\OC(n^2K^2)$ for testing the $K^2$ elements of $\GC$ once the Smith
form is available.  By using the group property of $\GC^{\pt}$ one can construct 
a faster algorithm, but that is beyond the scope of this paper.

\section{Correctable $*$-Algebra\label{apdxD}}

The counterpart in \cite{PhysRevA.76.042303} of our notion of information
perfectly present at the output of a quantum channel, see Sec.~\ref{sct2}, is
that of a \emph{correctable $*$-algebra} $\AC$ of operators acting on a
Hilbert space. The $*$ (sometimes denoted C$^*$) means that $\AC$, as well as
being an algebra of operators in the usual sense, contains $a\ad$ whenever
it contains $a$. Let the channel superoperator $\EC$ be represented by Kraus
operators,
\begin{equation}
\label{eqn95}
\EC(\rho) = \sum_j E_j\rho E_j\ad, 
\end{equation}
satisfying the usual closure condition $\sum_j E_j\ad E_j=I$, and let $P$ be a
projector onto some subspace $P\HC$ of the Hilbert space $\HC$.  Then a $*$-algebra $\AC$ is defined in \cite{PhysRevA.76.042303} to be \emph{correctable for $\EC$ on states in $P\HC$} provided $a=PaP$ for every $a$ in $\AC$, and there exists a superoperator $\RC$ (the recovery operation in an error correction scheme) whose domain is the range of $\EC$, whose range is $\LC(\HC)$, and such that
\begin{equation}
\label{eqn96}
P[(\RC \circ \EC)^\dagger (a)] P = a = P a P
\end{equation}
for all $a\in\AC$.  Here the dagger denotes the adjoint of the superoperator
in the sense that
\begin{equation}
\label{eqn97} 
\Tr\left[b \left((\RC\circ\EC)(c) \right)\right] = 
\Tr\left[ \left((\RC \circ \EC)^\dagger(b) \right) c\right]
\end{equation}
for any $b$ and $c$ in $\LC(\HC)$. 
In \cite{PhysRevA.76.042303}, see Theorem~9 and Corollary~10, it is shown that
any correctable algebra in this sense is a subalgebra of (what we call) a
\emph{maximal} correctable algebra
\begin{equation}
\label{eqn98} 
\AC_M = \left\{ a \in \LC({P\HC}) : 
[ a, P E^\dag_i E_j P]=0 \quad \forall \: i,j \right\}.
\end{equation}

We can apply this to our setting described in Secs.~\ref{sct4} and \ref{sct5} where $P$ is the projector on the coding space $\HC_C$ and $\EC_B$ is the superoperator for the partial trace down to the subset $\pt$ of carriers,
\begin{equation}
\label{eqn99} 
\EC_B(\rho) = \Tr_{\ptc}[\rho] = \sum_j E_j \rho E_j\ad \quad \text{ for }\rho \in \LC(\HC) 
\end{equation}
with Kraus operators
\begin{equation}
\label{eqn100}
E_j := I_B \otimes \bra{j}_{\ptc},
\end{equation}
where $\ket{j}_{\ptc}$ is any orthonormal basis of $\HC_{\ptc}$, so
\begin{equation}
\label{eqn101}
E_i\ad E_j = I_\pt \otimes \ket{i}\bra{j}_{\ptc}.
\end{equation}

We shall now show that collection of operators in $\GC^\pt$ (defined in Theorem~\ref{thm3}) spans a $*$-algebra which is correctable for $\EC_\pt$ on states in $P\HC = \HC_C$, and is the maximal algebra of this kind, i.e. $\text{span}(\GC^{\pt}) = \AC_M$.  First note that $\text{span}(\GC^{\pt})$ is indeed a $*$-algebra: every $\hat g \in \GC$ is a unitary operator and $\GC$ contains the adjoint of each of its elements; replacing $g$ with $g\ad$ in \eqref{eqn48} yields $\hat g\ad$. Of course $\Tr_{\ptc}[\hat g]=0$ if and only if $\Tr_{\ptc}[\hat g\ad]=0$ and in addition, $a = P a P$ for $a \in \text{span}(\GC^{\pt})$ because $\hat g = P\hat g P$, \eqref{eqn48}.

By definition $\Tr_{\ptc}[\hat g]\neq 0$ for $\hat g \in \GC^\pt$, and this means that the partial trace down to $\pt$ of at least one element in the corresponding coset $g\SC$, see \eqref{eqn48}, must be nonzero.  Let $h$ be such an element; since it is a Pauli product it must be of the form $h=h_\pt\otimes I_{\ptc}$.  As a consequence,
\begin{align}
\label{eqn102}
[\hat g,PE_i\ad E_j P] &=  [\hat h,PE_i\ad E_j P]= 
 P[h, E_i\ad E_j] P\notag\\
 &= P[\,h_\pt\otimes I_{\ptc}, I_\pt \otimes \ket{i}\bra{j}_{\ptc}\,] P = 0,
\end{align}
where the successive steps are justified as follows.  Since $\hat g$ depends
only on the coset $g\SC$ and $h$ belongs to this coset, $h\SC=g\SC$ and $\hat
h = Ph = hP =\hat g$.  This means we can move the projector $P$ outside the
commutator bracket, and once outside it is obvious that the latter vanishes
for every $i$ and $j$.   Thus any $\hat g$ in $\GC^\pt$ belongs
to the maximal $\AC_M$ defined in \eqref{eqn98}, as do all linear
combinations of the elements in $\GC^\pt$.

To show that $\AC_M$ is actually spanned by $\GC^\pt$ we note that any 
$a$ belonging to  $\AC_M$ can be written as  
\begin{equation}
\label{eqn103} 
a = b+c,
\end{equation}
where $b$ is a linear combination of elements of $\GC^\pt$ and $c$ of elements
of $\GC$ that do not belong to $\GC^\pt$, so
$\Tr_{\ptc}[c]=\Tr_{\ptc}[c\ad]=0$. Thus it is the case that
\begin{equation}
\label{eqn104} 
P(\RC\circ\EC_\pt)\ad (b)P = b,\quad 
P(\RC\circ\EC_\pt)\ad (c)P = c,
\end{equation}
where the first follows, see \eqref{eqn96}, from the previous argument showing
that the span of $\GC^\pt$ is a subalgebra of $\AC_M$, and the second from
linearity and the assumption that $a$ belongs to $\AC_M$.  Multiply the second
equation by $c\ad$ and take the trace:
\begin{align}
\label{eqn105}
 \Tr[c\ad c] &= \Tr \left[ c\ad P \left( (\RC\circ\EC_\pt)\ad (c) \right) P \right]\notag\\
  &= \Tr \left[ \left( \RC\circ\EC_\pt(c\ad) \right) c \right] = 0,
\end{align}
where we used the fact that $Pc\ad P=c\ad$, and $\EC_\pt(c\ad) = \Tr_{\ptc}[c\ad]=0$.  Thus $c=0$ and any element of $\AC_M$ is a linear combination of the operators in $\GC^\pt$.

In conclusion, we have shown for any additive graph code $C$ and any subset of
carrier qudits $\pt$, the $*$-algebra spanned by operators in $\GC^\pt$ is
exactly the maximal correctable algebra $\AC_M$ defined in
 \eqref{eqn98}. In App.~\ref{apdxC} we outline an algorithm that
enumerates the elements in $\GC^{\pt}$ for any $\HC_C$ and $\EC_\pt$, which in
light of the result above is an operator basis of $\AC_M$.



\begin{thebibliography}{10}

\bibitem{PhysRevA.51.2738}
Benjamin Schumacher.
\newblock Quantum coding.
\newblock {\em Phys. Rev. A}, 51(4):2738--2747, Apr 1995.

\bibitem{PhysRevA.52.R2493}
Peter~W. Shor.
\newblock Scheme for reducing decoherence in quantum computer memory.
\newblock {\em Phys. Rev. A}, 52(4):R2493--R2496, Oct 1995.

\bibitem{PhysRevLett.76.722}
Charles~H. Bennett, Gilles Brassard, Sandu Popescu, Benjamin Schumacher,
  John~A. Smolin, and William~K. Wootters.
\newblock Purification of noisy entanglement and faithful teleportation via
  noisy channels.
\newblock {\em Phys. Rev. Lett.}, 76(5):722--725, Jan 1996.

\bibitem{PhysRevA.54.1098}
A.~R. Calderbank and Peter~W. Shor.
\newblock Good quantum error-correcting codes exist.
\newblock {\em Phys. Rev. A}, 54(2):1098--1105, Aug 1996.

\bibitem{MacWilliams:Sloane}
F.~J. MacWilliams and N.~J.~A. Sloane.
\newblock {\em The Theory of Error-Correcting Codes}.
\newblock North-Holland Mathematical Library, Amsterdam, 1977.

\bibitem{PhysRevLett.77.198}
Raymond Laflamme, Cesar Miquel, Juan~Pablo Paz, and Wojciech~Hubert Zurek.
\newblock Perfect quantum error correcting code.
\newblock {\em Phys. Rev. Lett.}, 77(1):198--201, Jul 1996.

\bibitem{PhysRevA.71.042337}
Robert~B. Griffiths.
\newblock Channel kets, entangled states, and the location of quantum
  information.
\newblock {\em Phys. Rev. A}, 71(4):042337, 2005.

\bibitem{PhysRevA.78.042303}
Shiang~Yong Looi, Li~Yu, Vlad Gheorghiu, and Robert~B. Griffiths.
\newblock Quantum-error-correcting codes using qudit graph states.
\newblock {\em Phys. Rev. A}, 78(4):042303, 2008.

\bibitem{quantph.0111080}
Dirk Schlingemann.
\newblock Stabilizer codes can be realized as graph codes.
\newblock e-print arXiv:quant-ph/0111080.

\bibitem{quantph.0703112}
Markus Grassl, Andreas Klappenecker, and Martin Roetteler.
\newblock Graphs, quadratic forms and quantum codes.
\newblock e-print arXiv:quant-ph/0703112.

\bibitem{quantph.9802007}
Daniel Gottesman.
\newblock Fault-tolerant quantum computation with higher-dimensional systems.
\newblock {\em Chaos Solitons Fractals}, 10:1749, 1999.
\newblock e-print arXiv:quant-ph/9802007.

\bibitem{IEEETransInfTheory.45.1827}
Eric~M. Rains.
\newblock Nonbinary quantum codes.
\newblock {\em IEEE Trans. Inf. Theory}, 45(6):1827--1832, Sep 1999.

\bibitem{PhysRevA.65.012308}
D.~Schlingemann and R.~F. Werner.
\newblock Quantum error-correcting codes associated with graphs.
\newblock {\em Phys. Rev. A}, 65(1):012308, Dec 2001.

\bibitem{PhysRevA.78.012306}
Dan Hu, Weidong Tang, Meisheng Zhao, Qing Chen, Sixia Yu, and C.~H. Oh.
\newblock Graphical nonbinary quantum error-correcting codes.
\newblock {\em Physical Review A (Atomic, Molecular, and Optical Physics)},
  78(1):012306, 2008.

\bibitem{PhysRevA.75.032345}
Lev Ioffe and Marc M{\'e}zard.
\newblock Asymmetric quantum error-correcting codes.
\newblock {\em Phys. Rev. A}, 75(3):032345, 2007.

\bibitem{PhysRevLett.98.100502}
C{\'e}dric B{\'e}ny, Achim Kempf, and David~W. Kribs.
\newblock Generalization of quantum error correction via the heisenberg
  picture.
\newblock {\em Phys. Rev. Lett.}, 98(10):100502, 2007.

\bibitem{PhysRevA.76.042303}
C{\'e}dric B{\'e}ny, Achim Kempf, and David~W. Kribs.
\newblock Quantum error correction of observables.
\newblock {\em Phys. Rev. A}, 76(4):042303, 2007.

\bibitem{arxiv0907.4207}
C{\'e}dric B{\'e}ny.
\newblock Conditions for the approximate correction of algebras.
\newblock e-print arXiv:0907.4207 [quant-ph].

\bibitem{PhysRevLett.100.030501}
Robin Blume-Kohout, Hui~Khoon Ng, David Poulin, and Lorenza Viola.
\newblock Characterizing the structure of preserved information in quantum
  processes.
\newblock {\em Phys. Rev. Lett.}, 100(3):030501, 2008.

\bibitem{PhysRevA.76.062320}
Robert~B. Griffiths.
\newblock Types of quantum information.
\newblock {\em Phys. Rev. A}, 76(6):062320, 2007.

\bibitem{PhysRevA.54.2759}
Robert~B. Griffiths.
\newblock Consistent histories and quantum reasoning.
\newblock {\em Phys. Rev. A}, 54(4):2759--2774, Oct 1996.

\bibitem{RBGriffiths:ConsistentQuantumTheory}
Robert~B. Griffiths.
\newblock {\em Consistent Quantum Theory}.
\newblock Cambridge University Press, Cambridge, 2002.

\bibitem{PhysRevA.71.042315}
Erik Hostens, Jeroen Dehaene, and Bart~De Moor.
\newblock Stabilizer states and clifford operations for systems of arbitrary
  dimensions and modular arithmetic.
\newblock {\em Phys. Rev. A}, 71(4):042315, 2005.

\bibitem{Newman:IntegralMatrices}
Morris Newman.
\newblock {\em Integral Matrices}.
\newblock Academic Press, New York, 1972.

\bibitem{Storjohann96nearoptimal}
Arne Storjohann.
\newblock Near optimal algorithms for computing smith normal forms of integer
  matrices.
\newblock {\em ISSAC '96: Proceedings of the 1996 international symposium on Symbolic and algebraic computation}.
\newblock pages 267--274. ACM Press, 1996.

\bibitem{CoppersmithWinograd}
D.~Coppersmith and S.~Winograd.
\newblock Matrix multiplication via arithmetic progressions.
\newblock {\em STOC '87: Proceedings of the nineteenth annual ACM symposium
  on Theory of computing}, pages 1--6, New York, NY, USA, 1987. ACM.

\bibitem{NielsenChuang:QuantumComputation}
Michael~A. Nielsen and Isaac~L. Chuang.
\newblock {\em Quantum Computation and Quantum Information}.
\newblock Cambridge University Press, Cambridge, 5th edition, 2000.

\bibitem{PhysRevA.56.33}
M.~Grassl, Th. Beth, and T.~Pellizzari.
\newblock Codes for the quantum erasure channel.
\newblock {\em Phys. Rev. A}, 56(1):33--38, Jul 1997.

\bibitem{quantph.0709.1780}
Sixia Yu, Qing Chen, and C.H. Oh.
\newblock Graphical quantum error-correcting codes.
\newblock e-print arXiv:0709.1780 [quant-ph].

\bibitem{HornJohnson:MatrixAnalysis}
Roger~A. Horn and Charles~R. Johnson.
\newblock {\em Matrix Analysis}.
\newblock Cambridge University Press, Cambridge, 1999.

\end{thebibliography}

\end{document}